\documentclass[aps,preprint,showpacs,nofootinbib,showkeys,prd]{revtex4-2}

\usepackage{graphicx}  %
\usepackage{amsmath,amssymb} 
\usepackage{bm}  %
\usepackage{bbm}
\usepackage{ulem}  
\usepackage{multirow}
\usepackage{dsfont}
\usepackage{comment}
\usepackage[mathlines]{lineno}
\usepackage{xcolor}

\newcommand{\bea}{\begin{eqnarray}}
\newcommand{\eea}{\end{eqnarray}}
\newcommand{\beq}{\begin{equation}}
\newcommand{\eeq}{\end{equation}}
\newcommand{\bqa}{\begin{eqnarray}}
\newcommand{\eqa}{\end{eqnarray}}

\def\mqo2{{\!\!\!}}

\usepackage{relsize}
\def\babar{\mbox{\slshape B\kern-0.1em{\smaller A}\kern-0.1em
    B\kern-0.1em{\smaller A\kern-0.2em R}}}
    
\begin{document}

\preprint{TUM-EFT 188/24}
\title{Born-Oppenheimer Potentials for $\bm{SU(3)}$ Gauge Theory}
\author{Fareed Alasiri}
\email{alasiri.6@osu.edu}
\affiliation{Department of Physics,
        The Ohio State University, Columbus, OH\ 43210, USA}
\author{Eric Braaten}
\email{braaten.1@osu.edu}
\affiliation{Department of Physics,
         The Ohio State University, Columbus, OH\ 43210, USA}
\author{Abhishek Mohapatra}
\email{abhishek.mohapatra@tum.de}
\affiliation{Technical University of Munich,\\
TUM School of Natural Sciences, Physics Department,\\   
James-Franck-Str.~1, 85748 Garching, Germany.}\date{\today}

\begin{abstract}
We develop parameterizations of  8 of the lowest  Born-Oppenheimer potentials 
for quarkonium hybrid mesons
as functions of the separation $r$  of the static quark and antiquark sources.
The parameters are determined by fitting results calculated using pure $SU(3)$ lattice gauge theory.
The parameterizations have the correct limiting behavior at small $r$, 
where the potentials form multiplets associated with gluelumps.
They have the correct limiting behavior at large $r$, 
where the potentials form multiplets associated with excitations of a relativistic string.
There is a narrow avoided crossing in the small-$r$ region between two  potentials
with the same Born-Oppenheimer quantum numbers. 
\end{abstract}

\smallskip
\pacs{14.40.Pq,14.40.Rt,31.30.-i,13.25.Gv}
\keywords{
Born-Oppenheimer approximation, heavy quarks, quarkonium hybrid mesons.}
\maketitle

\newpage

\section{Introduction}
\label{sec:Intro}

The discovery of dozens of {\it exotic heavy hadrons} over the last two decades 
presents a serious challenge to our understanding of {\it quantum chromodynamics} (QCD) \cite{Brambilla:2019esw}.
Most previous efforts to understand the exotic heavy hadrons have been based on models 
that  make very little connection with the fundamental theory.
One approach to the problem that is firmly based on QCD is the {\it Born-Oppenheimer approximation}.
The Born-Oppenheimer (B\nobreakdash-O) approximation for QCD 
was pioneered by Juge, Kuti, and Morningstar (JKM) in 1999  \cite{Juge:1999ie}. 
It separates the problem into two steps:
(1) the calculation of B\nobreakdash-O potentials that give the energy levels of QCD 
in the presence of static color sources,
(2) the solution of the Schr\"odinger equation for heavy quarks and antiquarks in the B\nobreakdash-O potentials.
The connection to the fundamental theory is through the  calculation of B\nobreakdash-O potentials using lattice QCD
and through the mass of the heavy quark in the Schr\"odinger equation.
The B\nobreakdash-O approximation has been developed into an effective field theory 
that was originally named {\it potential NRQCD} \cite{Brambilla:1999xf} and is now
called {\it Born-Oppenheimer effective field theory} (BOEFT)
 \cite{Berwein:2015vca,Oncala:2017hop,Brambilla:2017uyf,Soto:2020xpm, Berwein:2024ztx}.
 BOEFT allows systematically improvable approximations to various properties of hadrons whose constituents include more than one heavy quark or antiquark.
It has been applied to quarkonium hybrids 
\cite{Brambilla:2018pyn,Brambilla:2019jfi,TarrusCastella:2021pld,Brambilla:2022hhi,Soto:2023lbh},
double-heavy baryons \cite{Soto:2020pfa},
and to conventional quarkonium \cite{TarrusCastella:2022rxb}.

The B\nobreakdash-O potentials for hadrons containing a heavy quark and antiquark
are the discrete energy levels of QCD
in the presence of static color-triplet ($\bm{3}$) and color-antitriplet ($\bm{3^*}$) sources
separated by a variable distance $r$.
In Ref.~\cite{Juge:1999ie}, JKM used lattice gauge theory to calculate the lowest 3
B\nobreakdash-O potentials for pure $SU(3)$ gauge theory.  
The bound states in the lowest potential  (labeled by B\nobreakdash-O quantum numbers $\Sigma_g^+$) 
were identified as quarkonium mesons.
The bound states in the next higher potentials 
(labeled by $\Pi_u$ and $\Sigma_u^-$) were identified as  quarkonium hybrid mesons.
In Ref.~\cite{Juge:2002br}, JKM extended their calculations to potentials with
many other B\nobreakdash-O quantum numbers.
The B\nobreakdash-O potentials for pure $SU(3)$ gauge theory
are all {\it confining potential}s that increase linearly at large $r$.
We refer to all the higher potentials after the $\Sigma_g^+$  quarkonium potential as {\it hybrid potentials}.
JKM pointed out that there is a rearrangement of the hybrid potentials between small $r$ and large $r$.
At small $r$, the hybrid potentials form multiplets associated with {\it gluelumps}, 
which are energy levels of QCD in the presence of a color-octet ($\bm{8}$) source.
At large $r$, the hybrid potentials form multiplets associated with excitations of a {\it relativistic string},
which can be interpreted as a model for the gluon flux tube extending between the $\bm{3}$ and $\bm{3^*}$ sources.
Morningstar has collected many B\nobreakdash-O potentials calculated by JKM on a website \cite{Morningstar}.
There have been several recent calculations of the  B\nobreakdash-O potentials for pure $SU(3)$ gauge theory
using more sophisticated methods \cite{Capitani:2018rox,Schlosser:2021wnr,Bicudo:2021tsc,Sharifian:2023idc}.
The goal of this paper is to develop simple parameterizations for some of the lower  B\nobreakdash-O potentials.

In QCD, the most dramatic effect of light quarks on B\nobreakdash-O potentials is the existence of new 
potentials that approach constants at large $r$.
The constants are the thresholds for pairs of {\it static hadrons},
which are QCD fields bound to a $\bm{3}$ or $\bm{3^*}$ source.
We refer to these B\nobreakdash-O potentials as {\it meson-pair potentials}.
The lowest $\Sigma_g^+$ potentials with isospin 0 have been calculated
for QCD with 2 and 3 flavors of light quarks  \cite{Bali:2005fu,Bulava:2019iut,Bulava:2024jpj}.  
They exhibit narrow avoided crossings 
between confining potentials and meson-pair potentials that can be attributed 
to the breaking of the gluon flux tube by the creation of a light quark-antiquark pair.
The $\Pi_u$ hybrid potential has been calculated for QCD with 2 flavors of light quarks. 
Near its minimum, it does not seem to differ much from the
corresponding potential in pure $SU(3)$ gauge theory \cite{Bali:2000vr}.  
The calculations of other hybrid  B\nobreakdash-O potentials for QCD with light quarks
are still in their infancy \cite{Hollwieser:2023bud}.

In Refs.~\cite{Braaten:2013boa,Braaten:2014qka}, it was pointed out 
that QCD also has B\nobreakdash-O potentials that correspond to energy levels
in sectors of QCD whose light-quark flavor
is not a singlet  with respect to the $SU(3)$ flavor symmetry relating the up, down, and strange quarks.
We refer to B\nobreakdash-O potentials whose flavor is that of a light quark and a light antiquark 
as {\it tetraquark potentials}.
The first tetraquark  B\nobreakdash-O potentials with isospin 1 
have been calculated using lattice QCD with two flavors of  light quarks \cite{Prelovsek:2019ywc,Sadl:2021bme}.

Until more calculations of B\nobreakdash-O potentials  using lattice QCD with light quarks
 are available, the best one can do is develop models for the potentials.
The existence of light quarks introduces new B\nobreakdash-O potentials
at small $r$ that are associated with {\it adjoint hadrons}, 
which are QCD fields with nonsinglet light-quark flavors bound to an $\bm{8}$ source.
 A plausible model for such a potential is the corresponding potential from pure $SU(3)$ gauge theory
 with the gluelump energy replaced by the energy of the adjoint meson.
 
As a first step toward developing models for the B\nobreakdash-O potentials for QCD with light quarks,
it is useful to have more accurate parameterizations of the B\nobreakdash-O potentials  
for pure $SU(3)$ gauge theory.
In Section~\ref{sec:BO-QCD}, we describe  theoretical constraints on the  B\nobreakdash-O potentials.
In Section~\ref{sec:LatticeQCD}, we describe the information  about the  B\nobreakdash-O potentials
of pure $SU(3)$ gauge theory that is presently available.
In Section~\ref{sec:SU(3)Potentials}, we present parameterizations of 8 of the lowest   B\nobreakdash-O potentials
of pure $SU(3)$ gauge theory.
In Section~\ref{sec:Summary}, we discuss how the parameterizations can be modified 
to take into account the effects of  light quarks in QCD.


\section{Born-Oppenheimer Potentials}
\label{sec:BO-QCD}

In this section, we describe theoretical constraints on the B\nobreakdash-O potentials 
for QCD and for pure $SU(3)$ gauge theory.

\subsection{Born-Oppenheimer quantum numbers}
\label{sec:BO-quantum}

QCD has a rotational symmetry group $SO(3)$ that is generated 
by the total angular momentum vector $\bm{J}$ of the QCD fields.
It also has two discrete symmetries:  parity  $\mathcal{P}$ and charge conjugation $\mathcal{C}$.
Because of these symmetries,  
the QCD  Hamiltonian commutes with $\bm{J}$, $\mathcal{P}$, and $\mathcal{C}$.
Its  eigenstates can be chosen to have definite 
angular-momentum quantum numbers $(J,M)$ and discrete quantum numbers  $P$ and $C$.
The quantum numbers are traditionally specified in the form $J^{PC}$.

The presence of  static $\bm{3}$ and $\bm{3^*}$ color sources along the axis $\hat{\bm{r}}$ 
breaks the rotational symmetry of QCD down to the cylindrical symmetry subgroup $SO(2)$
generated by $\bm{J} \!\cdot\! \hat{\bm{r}}$.
The sources break the $\mathcal{C}$ and $\mathcal{P}$ symmetries of  QCD 
down to the single discrete symmetry $\mathcal{CP}$.
There is also an additional discrete symmetry: a reflection $\mathcal{R}$ through any specific plane containing $\bm{\hat{r}}$. 
The eigenvalues $\lambda$ of $\bm{J} \!\cdot\! \hat{\bm{r}}$ must be integer or half-integer.
The eigenvalues $\eta$ of $\mathcal{CP}$ are $+1$ or $-1$.
The eigenvalues $\epsilon$ of $\mathcal{R}$ are $+1$ or $-1$.
We refer to $\lambda$,  $\eta$, and $\epsilon$ as {\it B\nobreakdash-O quantum numbers}.
The QCD Hamiltonian, $(\bm{J} \!\cdot\! \hat{\bm{r}})^2$,
$\mathcal{C}\mathcal{P}$, and $\mathcal{R}$ can be simultaneously diagonalized.
Thus $|\lambda|$, $\eta$, and $\epsilon$ are independent B\nobreakdash-O quantum numbers.

The B\nobreakdash-O quantum numbers $|\lambda|$,  $\eta$, and $\epsilon$
are traditionally expressed in the form $\Lambda_\eta^\epsilon$, where $\Lambda$ specifies  
$|\lambda|$ according to the code $\Lambda=\Sigma,\Pi,\Delta, \ldots$ for $|\lambda| = 0,1,2, \ldots$,
the subscript $\eta$ is $g$ or $u$ if the $\mathcal{C}\mathcal{P}$  eigenvalue  is $+1$ or $-1$,
and the superscript $\epsilon$ is $+$ or $-$ if the $\mathcal{R}$ eigenvalue is  $+1$ or $-1$.
If $\Lambda \neq \Sigma$, the superscript $\epsilon$ is often omitted 
because the cylindrical symmetry requires $\Lambda_\eta^+$ and $\Lambda_\eta^-$ states 
to be degenerate in energy.
The ground state with B\nobreakdash-O quantum numbers $\Lambda_\eta^\epsilon$
 is labeled simply by $\Lambda_\eta^\epsilon$.
The excited states with the same quantum numbers are labeled 
$\Lambda_\eta^{\epsilon\prime}$, $\Lambda_\eta^{\epsilon\prime\prime}$, \ldots.

In pure $SU(3)$ gauge theory,
the lowest B\nobreakdash-O potentials are the ground-state $\Sigma_g^+$ quarkonium potential
and then the  $\Pi_u$ hybrid potential.
The next higher hybrid potentials at small $r$ are $\Sigma_u^-$, $\Sigma_g^{+ \prime}$, and $\Pi_g$.
The next higher hybrid potentials at large $r$ are $\Sigma_g^{+ \prime}$, $\Delta_g$, and $\Pi_g$. 
Potentials with different B\nobreakdash-O quantum numbers can cross as functions of $r$,
but there can be no crossings between potentials with the same B\nobreakdash-O quantum numbers.

\subsection{Short-distance behavior}
\label{sec:Vshort}

An effective field theory for heavy quarkonium called {\it potential  NRQCD}
(pNRQCD) \cite{Brambilla:1999xf}
can be used to provide rigorous constraints on the ground-state  $\Sigma_g^+$ potential at small $r$.
The potential can be expressed as the sum of a perturbative QCD potential 
for a color-singlet $Q \bar Q$ pair and nonperturbative corrections  \cite{Brambilla:1999xf}. 
The  perturbative QCD potential can be expanded in powers of the running coupling constant $\alpha_s(1/r)$
at the scale $1/r$.
The leading term is $-(4/3) \alpha_s(1/r)/r$.
The nonperturbative corrections can be expanded in integer powers of $r$.
The leading terms without any suppression factors of $\alpha_s(1/r)$ are $r^2$ and then $r^4$.

In our parametrization of the $\Sigma_g^+$ potential,
we will approximate the perturbative QCD potential by $\kappa_1/r$, 
where $\kappa_1$ is an adjustable parameter.
At small $r$, we will require  the ground-state  $\Sigma_g^+$ potential to have the form
\beq
V_{\Sigma_g^+}(r) = \frac{\kappa_1}{r} + E_{0^{++}} +  A_{\Sigma_g^+} \, r^2 + \mathcal{O}(r^4),
\label{VSigmag+-smallr}
\eeq
where the constants $E_{0^{++}}$ and $A_{\Sigma_g^+}$ are also adjustable parameters.

 In Ref.~\cite{Juge:2002br}, JKM  
pointed out that the hybrid B\nobreakdash-O  potentials  at small $r$ form multiplets
that become degenerate in the limit $r \to 0$.
A multiplet consists of $J+1$ potentials that correspond to different spin states 
of a gluelump with specific $J^{PC}$ quantum numbers.
The multiplet associated with a  $J^{PC}$ gluelump consists of potentials  $\Lambda_\eta^\epsilon$ with $\eta = CP$ 
and $\Lambda$ having all the values corresponding to $|\lambda| = 0, \ldots, J$.
For the $\Sigma_\eta^\epsilon$ member of the multiplet, the reflection quantum number is $\epsilon = P (-1)^J$.
The lowest multiplet is associated with the $1^{+-}$ gluelump and consists of $\Pi_u$  and $\Sigma_u^-$.
The next higher multiplet is associated with the $1^{--}$ gluelump and consists of $\Sigma_g^{+\prime}$ and $\Pi_g$.
The next higher multiplet is associated with the $2^{--}$ gluelump and consists of  $\Delta_g$, $\Pi_g^\prime$, and $\Sigma_g^-$.

Potential NRQCD has been extended into a more general effective field theory called  BOEFT
that can also be applied to quarkonium hybrids and other exotic heavy hadrons
 \cite{Berwein:2015vca,Oncala:2017hop,Brambilla:2017uyf,Soto:2020xpm}.
BOEFT can be used to deduce the asymptotic behavior of a hybrid B\nobreakdash-O potential at small $r$.
It can be expressed as the sum of a perturbative QCD potential for a color-octet $Q \bar Q$ pair 
and nonperturbative corrections  \cite{Berwein:2015vca,Brambilla:1999xf}. 
The perturbative QCD potential can be expanded in powers 
of the running coupling constant $\alpha_s(1/r)$ at the scale $1/r$.
The leading term  is $+(1/6) \alpha_s(1/r)/r$.
The nonperturbative corrections can be expanded in integer powers of $r$.  
The leading terms without any suppression factors of $\alpha_s(1/r)$ are $r^2$ and then $r^4$.

In our parameterizations of hybrid  B\nobreakdash-O potentials,
we will approximate the perturbative QCD potential by $\kappa_8/r$, where $\kappa_8$ is an adjustable parameter.
At small $r$, we will require a hybrid potential with B\nobreakdash-O quantum numbers 
$\Lambda_\eta^\epsilon$ in the multiplet of the $J^{PC}$ gluelump to have the form
\beq
V_{\Lambda_\eta^\epsilon}(r) \longrightarrow \frac{\kappa_8}{r} + E_{J^{PC}}  
+ A_{\Lambda_\eta^\epsilon} \,r^2  + \mathcal{O}(r^4),
\label{VLambda-smallr}
\eeq
where the constants $E_{J^{PC}}$ and  $A_{\Lambda_\eta^\epsilon}$ are also adjustable parameters.
The constant $\kappa_8$ is the same for all hybrid potentials.
The constant $E_{J^{PC}}$, which can be identified with the energy of the $J^{PC}$ gluelump,
is the same for all the $J+1$ potentials in the multiplet. 
The constants $A_{\Lambda_\eta^\epsilon}$ and $B_{\Lambda_\eta^\epsilon}$ are different for each of the $J+1$ potentials.

\subsection{Long-distance behavior}
\label{sec:Vlong}

In Ref.~\cite{Juge:2002br}, JKM pointed out that the B\nobreakdash-O  potentials also form multiplets at large $r$. 
All the B\nobreakdash-O potentials for pure $SU(3)$ gauge theory increase approximately linearly at large $r$. 
The differences between potentials in different multiplets decrease as $1/r$,
but the differences  between potentials within a multiplet decrease as a higher power of $1/r$.
JKM found that the multiplets are consistent with the excitation levels of a relativistic string \cite{Juge:2002br}.
The string potential is
\beq
V_N(r) = \sqrt{\sigma^2\, r^2 + 2 \pi \big(N - \tfrac{1}{12}\big) \sigma},
\label{V-N}
\eeq
where $\sigma$ is the string tension and $N$ is the quantum number for transverse vibrational excitations of the string.
For $N=0$, this potential becomes complex for $r < \sqrt{\pi/(6\sigma)}$,
so it can be a good approximation only when $r$ is significantly larger than  $\sigma^{-1/2}$.
The integer $N$ is even if $\eta=g$ and odd if $\eta = u$.
The B\nobreakdash-O quantum numbers for the excitation  of the relativistic string up to level $N=4$ 
are listed in Ref.~\cite{Juge:2003ge}.
The $N=0$ level is $\Sigma_g^+$.
The $N=1$ level is $\Pi_u$.
The $N=2$ level consists of $\Sigma_g^{+\prime}$, $\Pi_g$, and $\Delta_g$.
The $N=3$ level consists of 6 B\nobreakdash-O quantum numbers including $\Sigma_u^-$.
The $N=4$ level consists of 12 B\nobreakdash-O quantum numbers including $\Sigma_g^-$ and $\Pi_g^\prime$.

The expansion of the  string potential in Eq.~\eqref{V-N} in powers of $1/r$ 
at large $r$ gives terms with all odd powers of $1/r$.
Luscher and Weisz have pointed out that the $1/r$ terms are universal in the sense that 
their coefficients receive no corrections from string self-interactions \cite{Luscher:2004ib}.
String self-interactions can give corrections to the string potentials in Eq.~\eqref{V-N} beginning at order $1/r^3$.
String rigidity can give corrections at order $1/r^4$ \cite{Braaten:1986bz,Braaten:1987gq}. 
The effects of the boundary of the string can also give corrections  at order $1/r^4$ \cite{Aharony:2010cx}.
Aharony and Field have shown that Lorentz invariance requires the cancellation 
of the various contributions to the coefficient of the $1/r^3$ term  \cite{Aharony:2010cx}.
The leading nonuniversal corrections to the string potentials in Eq.~\eqref{V-N}
are therefore proportional to $1/r^4$.

In our parameterizations of B\nobreakdash-O potentials
at large $r$, we will require a potential with quantum numbers 
$\Lambda_\eta^\epsilon$ and string excitation level $N$ to have the form
\beq
V_{\Lambda_\eta^\epsilon}(r)  =
V_N(r) + E_0 + \mathcal{O}(1/r^4),
\label{V-larger}
\eeq
where $V_N(r)$ is the string potential in Eq.~\eqref{V-N} and $E_0$ is an adjustable constant.
The energy $E_0$ is the same for the ground-state $\Sigma_g^+$ potential and all the hybrid potentials.


\section{Lattice QCD Inputs}
\label{sec:LatticeQCD}

In this section, we summarize results from lattice gauge theory relevant to 
the B\nobreakdash-O potentials for pure $SU(3)$ gauge theory.

\subsection{Gluelumps}
\label{sec:Gluelump}

As the separation $r$ of the $\bm{3}$ and $\bm{3^*}$ sources decreases to 0,
their effect on the fields of QCD becomes more and more like 
that of a linear combination of  a local color-singlet ({\bf 1}) source and a local color-octet ({\bf 8}) source.
Since a local color-singlet operator is no source at all, only the color-octet component acts as a source of QCD fields.
The rotational symmetry and the discrete symmetries of QCD are restored in the limit $r \to 0$.
The stationary states of QCD created by the color-octet component of the source
therefore approach states with definite $J^{PC}$ quantum numbers and with specific light-quark flavors.
The spectrum of QCD in the presence of a static color-octet source
can be calculated using  lattice QCD up to an additive constant.
In pure $SU(3)$ gauge theory, a state with discrete energy bound to an {\bf 8} source is called a {\it gluelump}. 

The first calculations of the spectrum of QCD in the presence of  an {\bf 8} source
were made by Foster and Michael  in 1998 \cite{Foster:1998wu}.
They calculated the energy differences for the lowest-energy gluelumps using pure $SU(3)$ gauge theory. 
The ground-state gluelump has $J^{PC}= 1^{+-}.$
The first two excited gluelumps have quantum numbers $1^{--}$ and $2^{--}$.
The gluelump energies for  pure $SU(3)$ gauge theory
have recently been calculated more precisely by Herr, Schlosser, and Wagner \cite{Herr:2023xwg}.
Their energies relative to the ground-state gluelump are 342(36) and 523(11)~MeV. The gluelumps with the next higher energies  are $2^{+-}$ and $0^{++}$
with relative energies  925(42) and 979(36)~MeV.

\subsection{Lattice gauge theory scales}
\label{sec:scales}

The calculation of  a B\nobreakdash-O potential using lattice gauge theory with lattice spacing $a$
produces a dimensionless potential $V(r)\,a$ as a function of a dimensionless radial variable $r/a$.
The dependence on the lattice spacing must then be eliminated in favor of a physical scale.
A common choice for the physical scale is the {\it Sommer scale} $r_0$,
which is defined in terms of the ground-state $\Sigma_g^+$ potential as \cite{Sommer:1993ce}
\beq
\frac{dV_{\Sigma_g^+}(r)}{d(1/r)} = -1.65 \quad \mathrm{at~} r = r_0.
\label{r-Sommer}
\eeq
In Ref.~\cite{Capitani:2018rox}, the B\nobreakdash-O potentials 
were presented in the form of a table of $V(r)\,a$ as functions of  $r/a$.
In Ref.~\cite{Schlosser:2021wnr}, improved B\nobreakdash-O potentials 
in which the source self-energies and some lattice discretization errors were subtracted
were presented in the form of a table of $V(r)$ in units of GeV as functions of $r$ in units of fm.
The physical units were obtained by choosing the Sommer scale to be $r_0 = 0.50$~fm.
This value can be used to reexpress the potentials in the form of $V(r)\, r_0$  as functions of $r/r_0$.

A great advantage of the  Sommer scale $r_0$  is that it can be determined with  high accuracy.
Another advantage of $r_0$ is that it can be determined using only the potential $V(r)$ near $r=r_0$.
If the $\Sigma_g^+$ potential  at the data points of an ensemble near $r_0$ 
is approximated by the Cornell potential,
\beq
V_{\Sigma_g^+}(r) \approx \frac{\kappa_0}{r} + E_0 + \sigma_0\, r,
\label{V-Cornell}
\eeq
the Sommer scale can be approximated by
\beq
r_0 \approx \sqrt{\frac{1.65+\kappa_0}{\sigma_0}} .
\label{r0-Cornell}
\eeq
A more accurate estimate for the Sommer scale can be obtained by 
approximating the potential by a smooth interpolation of all the data points  in the ensemble
and then solving Eq.~\eqref{r-Sommer} for $r_0$ using the interpolating function.
This estimate  can be illustrated by applying it 
to the $\Sigma_g^+$ potential from the HYP2 ensemble in Ref.~\cite{Schlosser:2021wnr}.
The potential  is expressed in terms of physical units 
by setting $r_0 = 0.50$~fm.
The estimate for the Sommer scale  obtained using a smooth interpolating function 
determined by the 11 points in the potential is $r_0 \approx 0.502(12)$~fm.

A choice for the physical scale that has a simpler physical interpretation is the {\it string tension} $\sigma$,
which is defined by requiring the asymptotic behavior of the ground-state $\Sigma_g^+$ potential to be
\beq
V_{\Sigma_g^+}(r) \longrightarrow \sigma\, r  + V_0^e \quad \mathrm{as~} r \to \infty,
\label{r-sigma}
\eeq
where $V_0^e$ is a constant that depends on the lattice gauge theory ensemble $e$.
The string tension can be interpreted as the energy per length of a gluon flux tube
connecting $\bm{3}$ and $\bm{3^\ast}$ sources.
The B\nobreakdash-O potentials in Refs.~\cite{Morningstar,Bicudo:2021tsc,Sharifian:2023idc}
are presented in the form of dimensionless potentials  $V(r)/\sqrt{\sigma}$
as functions of a dimensionless radial variable $r\sqrt{\sigma}$.
The value of the string tension was given in Ref.~\cite{Sharifian:2023idc} as $\sigma \approx 0.18~\mathrm{GeV}^2$, 
but that value was never used.

The string tension $\sigma$ for an ensemble can only be determined by a numerical fit of the $\Sigma_g^+$ potential 
at the largest values of $r$ available. 
The accuracy to which $\sigma$ can be determined is much smaller than that for $r_0$.
The fit that determines $\sigma$ introduces a systematic error that is difficult to control.
We emphasize the dependence of the systematic error in the string tension 
on the  ensemble by denoting it by $\sigma_e$,
where the subscript $e$ specifies the ensemble.

\subsection{Born-Oppenheimer Potentials}
\label{sec:Hybrid}

Calculations of the ground-state $\Sigma_g^+$ quarkonium potential go back to the early days of lattice gauge theory.
The first calculations of the ground-state $\Pi_u$ and $\Sigma_u^-$  potentials in pure $SU(3)$ gauge theory
were by Juge, Kuti, and Morningstar (JKM)  in 1999  \cite{Juge:1999ie}.
In Ref.~\cite{Juge:2002br}, they extended their calculations to the ground-state potentials 
for the other 5 B\nobreakdash-O quantum numbers of the form $\Sigma^\epsilon_\eta$, $\Pi_\eta$, and $\Delta_\eta$
as well as the first excited-state potential $\Sigma_g^{+\prime}$. 
The values of $r$ ranged from 0.2 to 2.4~fm.
Morningstar's website gives 16   B\nobreakdash-O potentials 
with quantum numbers of the form $\Sigma^\epsilon_\eta$, $\Pi_\eta$, and $\Delta_\eta$.
If we set $\sigma = 0.18 ~\mathrm{GeV}^2$,  the values of $r$ range from 0.13 to 1.95~fm \cite{Morningstar}.  In addition to the ground-state potentials,
they include the first 3 excited-state potentials for  $\Pi_u$, 
the first 2  excited-state potentials for $\Sigma_g^+$ and $\Pi_g$,
and the first excited-state potential for  $\Delta_g$.

Capitani et al.\  have carried out precise calculations of the 8 ground-state potentials 
with B\nobreakdash-O quantum numbers $\Sigma^\epsilon_\eta$, $\Pi_\eta$, and $\Delta_\eta$
as well as the first excited-state potential $\Sigma_g^{+\prime}$  \cite{Capitani:2018rox}. 
The  potentials were calculated using a single  lattice ensemble labeled HYP2
with $r$ ranging from about 0.19 to  1.12~fm. 
Schlosser and Wagner have extended the calculations of the ground-state 
$\Sigma_g^+$, $\Pi_u$, and $\Sigma_u^-$  potentials  to smaller $r$ \cite{Schlosser:2021wnr}.
They presented potentials from 4 different lattice ensembles labeled A, B, C, and D
with $r$ ranging from about 0.08 to 0.56~fm. 
The ensembles A, B, C, and D had successively smaller lattice spacings, 
allowing access to successively smaller values of $r$.
The ensemble HYP2 had the same lattice spacing as A, 
but it used smeared temporal gauge links to suppress contributions from high-energy states.

Bicudo, Cardoso and Sharifian have calculated many  excited-state $\Sigma_g^+$ potentials 
for pure $SU(3)$ gauge theory \cite{Bicudo:2021tsc}.
In addition to the ground-state $\Sigma_g^+$ potential, they
calculated 8 excited-state $\Sigma_g^+$ potentials using an ensemble labeled $S_4$
and 6 excited-state $\Sigma_g^+$ potentials using ensembles labeled $W_1$, $W_2$, and $W_4$.
The ensembles $W_1$, $W_2$, and $W_4$ had successively larger anisotropies 
and successively smaller temporal lattice spacings.
The ensembles $S_4$ had the same bare anisotropy as $W_4$ but it used an improved anisotropic action.
If we set $\sigma = 0.18 ~\mathrm{GeV}^2$, the values of $r$ for the $S_4$ ensemble range from  0.14 to 1.70~fm.
The values of $r$ for the $W_1$, $W_2$, and $W_4$ ensembles ranged from about 0.14 to 2.78~fm.
Sharifian, Cardoso, and Bicudo have extended the calculation using the $S_4$ ensemble to the potentials 
with the other 7  B\nobreakdash-O quantum numbers 
$\Sigma^\epsilon_\eta$, $\Pi_\eta$, and $\Delta_\eta$ \cite{Sharifian:2023idc}. 
In addition to the ground-state potentials, 
they calculated the first 6 excited-state potentials for $\Sigma_g^-$, $\Pi_u$, $\Pi_g$, and $\Delta_g$
and the first 5, 4, and 3 excited-state potentials for $\Sigma_u^+$, $\Delta_u$, and  $\Sigma_u^-$, respectively.

\subsection{Lattice discretization errors}
\label{sec:discretization}

The direct result of a lattice gauge theory calculation of a B\nobreakdash-O potential
is a dimensionless ``bare'' potential $V(r)\, a$ as a function of $r/a$,
where $a$ is the lattice spacing.
In order to obtain a finite result in the limit $a \to 0$, it is necessary to subtract the constant
self-energies of the $\bm{3}$ and $\bm{3^\ast}$ sources.
The accuracy of the calculation  can be improved by also subtracting lattice discretization errors.

In Ref.~\cite{Capitani:2018rox}, the  ``bare'' B\nobreakdash-O potentials $V_{\Lambda_\eta^\epsilon}^\mathrm{HYP2}(r)$  
for 9 quantum numbers $\Lambda_\eta^\epsilon$ were calculated using an ensemble labeled HYP2.
In Ref.~\cite{Schlosser:2021wnr}, the  bare B\nobreakdash-O potentials 
for $\Lambda_\eta^\epsilon =  \Sigma_g^+, \Pi_u,\Sigma_u^-$
were also calculated using 4 additional ensembles  $e$ labeled A, B, C, and D. 
Improved potentials $\tilde{V}_{\Lambda_\eta^\epsilon}^e(r)$
for $\Lambda_\eta^\epsilon =  \Sigma_g^+, \Pi_u,\Sigma_u^-$ in  the ensembles A, B, C, D,  and HYP2
were presented  in the form of $V(r)$ in GeV as functions of  $r$ in fm.
They can be expressed in the form of $V(r)\, r_0$ as functions of  $r/r_0$ 
by using the assumed value $r_0 = 0.5$~fm.

The improved potential  $\tilde{V}_{\Lambda_\eta^\epsilon}^e(r)$ 
for an ensemble $e$ in Ref.~\cite{Schlosser:2021wnr}
differs from the bare potential $V_{\Lambda_\eta^\epsilon}^e(r)$
 by the subtractions of the total self energy $C^e$  of the two sources,
a tree-level  lattice discretization error $\Delta V_{\Lambda_\eta^\epsilon}^e(r)$ 
that depends on $r$,
and a constant order-$a^2$  lattice discretization error $A_{\Lambda_\eta^\epsilon}^e\, a^2$.
In the case of the ground-state $\Sigma_g^+$ potential, 
the order-$a^2$  lattice discretization error can be absorbed into the self energy.
The relation between the bare and improved potentials is therefore
\beq
\tilde{V}_{\Sigma_g^+}^e(r) 
= V_{\Sigma_g^+}^e(r) - C^e - \Delta V_{\Sigma_g^+}^e(r).
\label{VSigmag+-imp}
\eeq
The self-energy $C^e$ depends on the ensemble.
The tree-level  lattice discretization errors $\Delta V_{\Sigma_g^+}^e(r)$ were expressed as  
functions of $r/a$ that depend on the ensemble $e$ multiplied by $\alpha^\prime/r$,
where $\alpha^\prime$ is the same  constant for all ensembles. 
The tree-level  lattice discretization errors are largest at  small $r$.

In the case of a hybrid potential $\Lambda_\eta^\epsilon$, 
the relation between the bare and improved potentials is
\beq
\tilde{V}_{\Lambda_\eta^\epsilon}^e(r) 
= V_{\Lambda_\eta^\epsilon}^e(r) - C^e + \tfrac18 \Delta V_{\Sigma_g^+}^e(r) - A_{\Lambda_\eta^\epsilon}^{\prime e}\, a^2.
\label{VLambda-imp}
\eeq
The  tree-level  lattice discretization error $\Delta V_{\Sigma_g^+}^e(r)$
differs from that for $\Sigma_g^+$ in Eq.~\eqref{VSigmag+-imp} by the multiplicative factor $-1/8$.
The coefficients $A_{\Lambda_\eta^\epsilon}^{\prime e}$
in the order-$a^2$  lattice discretization error 
were determined for $\Lambda_\eta^\epsilon =  \Pi_u$ and $\Sigma_u^-$. 
They are the same for the ensembles A, B, C, and D, but different for HYP2.

Having subtracted lattice discretization errors and expressed their potentials in terms of the Sommer scale $r_0$,
the improved potentials from the 5 ensembles in Ref.~\cite{Schlosser:2021wnr} all line up nicely
for each of the quantum numbers $\Sigma_g^+$, $\Pi_u$, and $\Sigma_u^-$.
In Refs.~\cite{Morningstar,Capitani:2018rox,Bicudo:2021tsc,Sharifian:2023idc},
lattice discretization errors were not subtracted.
One should therefore expect a spread in the potentials from the different ensembles at the smallest values of  $r$.
This problem can be avoided by taking the differences between hybrid potentials,
in which case the self energy and the tree-level lattice discretization energies in Eq.~\eqref{VLambda-imp} cancel.
The order-$a^2$  lattice discretization errors  in Eq.~\eqref{VLambda-imp} do not necessarily cancel.

\subsection{Transforming Potentials}
\label{sec:Potentials}

In Refs.~\cite{Morningstar,Bicudo:2021tsc,Sharifian:2023idc},  the potentials were expressed in terms 
of the string tension $\sigma_e$ instead of the Sommer scale $r_0$.
The systematic  error from the fitting of $\sigma_e$ can produce a spread in the energy offset 
between different ensembles and also a spread in the slope. 

Transforming  potentials that depend on $\sigma_e$
into potentials  that depend on $r_0$ is not entirely trivial.
The transformation of the potentials of Refs.~\cite{Morningstar,Bicudo:2021tsc,Sharifian:2023idc}
that allows them to be compared to the potentials of Refs~\cite{Capitani:2018rox,Schlosser:2021wnr}
involves the rescaling of $V(r)/\sqrt{\sigma_e}$ by a factor of $r_0 \sqrt{\sigma_e}$,
the rescaling of $r \sqrt{\sigma_e}$ by a factor of $1/\big(r_0 \sqrt{\sigma_e} \, \big)$,
and also  a shift in the energy by an additive constant.
The additive constants  in the potentials of Refs.~\cite{Capitani:2018rox,Schlosser:2021wnr} 
were essentially determined by fitting
the ground-state $\Sigma_g^+$ potential in the region $r > 0.2$~fm to the Cornell potential 
in Eq.~\eqref{V-Cornell} and then subtracting the constant $E_0$ from all the potentials.
The additive constants  in the potentials of Refs.~\cite{Morningstar,Bicudo:2021tsc,Sharifian:2023idc}
were determined by fitting the ground-state $\Sigma_g^+$ potential in the large-$r$ region
 to the linear potential $\sigma\, r  + V_0^e$ 
 and then subtracting the constant $V_0^e$ from all the potentials.

In order to transform the potentials of Refs.~\cite{Morningstar,Bicudo:2021tsc,Sharifian:2023idc}
into potentials that depend on $r_0$ instead of $\sigma$,
we exploit the fact that the $\Sigma_g^+$ potential over a small enough range of $r$ 
can be approximated by the Cornell potential in Eq.~\eqref{V-Cornell}.
The range  $\tfrac12 r_0 < r <2 r_0$ is narrow enough that the Cornell potential is a good approximation.
The transformation of the potential can be carried out using a sequence of 4 steps.

Step 1:
For all 5 ensembles in Refs.~\cite{Capitani:2018rox,Schlosser:2021wnr}, 
the dimensionless $\Sigma_g^+$ potential in the range $\tfrac12 r_0 < r < 2 r_0$
can be approximated by 
\beq
V_{\Sigma_g^+}(r)\,  r_0 \approx \frac{\kappa_0}{r/r_0} + r_0E_0 + (r_0^2\sigma_0) (r/r_0),
\label{Cornell-Wagner}
\eeq
with the same dimensionless coefficients $\kappa_0$, $E_0  r_0$, and $r_0^2\sigma_0$ 
for the HYP2, A, B, C, and D ensembles.
The coefficients can be determined by a 3-parameter  fit to the 16 points of the $\Sigma_g^+$ potentials  in the range $\tfrac12 r_0 < r < 2 r_0$ from the 5 ensembles.
The  fitted coefficients are 
\beq
\kappa_0 = -0.289(13), \quad r_0E_0 = - 0.003(29), \quad r_0^2\sigma_0  = 1.350(15). 
\label{Cornell-SW}
\eeq
The errors come only from the statistical errors in the dimensionless potentials $V_{\Sigma_g^+}(r)\,  r_0$.
Note that the fitted value of $r_0E_0$ is consistent with 0.

Step 2:  The dimensionless Sommer scale $r_0 \sqrt{\sigma_e}$ for an ensemble $e$  
can be estimated by approximating the potential by a smooth interpolation of all the data points  
and then solving Eq.~\eqref{r-Sommer} using the interpolating function.
The results for $r_0 \sqrt{\sigma_e}$ for each of the 4 ensembles in Ref.~\cite{Morningstar} 
and each of the 4 ensembles in  Ref.~\cite{Bicudo:2021tsc} are given in Table~\ref{tab:Cornellcoeffs}.

Step 3:
For any specific ensemble $e$ in 
Ref.~\cite{Morningstar} and  Ref.~\cite{Bicudo:2021tsc},
the dimensionless $\Sigma_g^+$ potential in the range $\tfrac12 r_0 < r < 2 r_0$
can be approximated by  
\beq
V_{\Sigma_g^+}^e(r)\big/\sqrt{\sigma_e}
 \approx \frac{\kappa_0}{r \sqrt{\sigma_e}} + \frac{E_0^\prime}{\sqrt{\sigma_e}}
+ \frac{\sigma_0}{\sigma_e} (r \sqrt{\sigma_e}).
\label{Cornell-Bicudo}
\eeq
Since $-1/r$ and $r$ are both increasing functions of $r$, a 3-parameter fit to the Cornell potential  in Eq.~\eqref{Cornell-Bicudo} 
gives values of $\kappa_0$ and $\sigma_0/\sigma_e$ that are strongly correlated.
We avoid this problem by choosing  their values so that the potential $V_{\Sigma_g^+}^e(r)$ from Eq.~\eqref{Cornell-Bicudo}
has the same dependence on $r$ as the potential $V_{\Sigma_g^+}(r)$ from Eq.~\eqref{Cornell-Wagner}.
The coefficient $\kappa_0$ in  Eq.~\eqref{Cornell-Bicudo} must  have the same value as  in Eq.~\eqref{Cornell-Wagner},
which is given in Eq.~\eqref{Cornell-SW}.
The coefficient $\sigma_0/\sigma_e$ in  Eq.~\eqref{Cornell-Bicudo} 
must differ from  $r_0^2 \sigma_0$ in Eq.~\eqref{Cornell-Wagner} by a factor of $1/(r_0^2 \sigma_e)$.
Given the estimate for $r_0 \sqrt{\sigma_e}$ from Step 2, 
the coefficient $\sigma_0/\sigma_e$
in Eq.~\eqref{Cornell-Bicudo} can be determined using Eq.~\eqref{r0-Cornell}:
\beq
\frac{\sigma_0}{\sigma_e}
 \approx \frac{1.65+ \kappa_0}{(r_0 \sqrt{\sigma_e})^2}.
\label{sigma0/sigma}
\eeq
The results for $\sigma_0/\sigma_e$ 
for  each of the 8 ensembles in Ref.~\cite{Morningstar} and Ref.~\cite{Bicudo:2021tsc} are given in Table~\ref{tab:Cornellcoeffs}.
Given the values of $\kappa_0$ and $\sigma_0/\sigma_e$,
we determine the energy offset $E_0^\prime/\sqrt{\sigma_e}$ in Eq.~\eqref{Cornell-Bicudo}
by a 1-parameter fit to the $\Sigma_g^+$ potential  in the range $\tfrac12 r_0 < r < 2 r_0$.
The results for $E_0^\prime/\sqrt{\sigma_e}$ for 
each of the 8 ensembles in Ref.~\cite{Morningstar} and Ref.~\cite{Bicudo:2021tsc} are given in Table~\ref{tab:Cornellcoeffs}.

\begin{table}[t]
\begin{tabular}{cc|cccc}
~References~ & ~ensemble~ & ~~~~~$r_0\sqrt{\sigma_e}$~~~~~ & ~~~~~$\kappa_0$~~~~~ 
& ~~~~~$\sigma_0/\sigma_e$~~~~~ & ~~$E'_0/\sqrt{\sigma_e}$ \\
\hline
\multirow{4}{*}{\cite{Morningstar}} 
& $A$  &  1.22(14) & $-0.289$ & 0.92(21) & ~~~~0.297(17)  \\
& $B$  &  1.20(29) & $-0.289$ & 0.95(46) & ~~~~0.211(15)  \\
& $C$  &  1.168(20) & $-0.289$ & 0.997(33) & ~~~~0.162(5) ~  \\
& $D$  & 1.153(20) & $-0.289$ & 1.024(35) & ~~~~0.106(11)  \\
\hline
\multirow{4}{*}{\cite{Bicudo:2021tsc}} 
&  $W_1$ & 1.150(15) & $-0.289$ & 1.030(26) & $-$0.0678(8)  \\
&  $W_2$ & 1.198(11) & $-0.289$ & 0.949(17) &  ~~~~0.2284(17)  \\
&  $W_4$ & 1.186(12) & $-0.289$ & 0.969(20) &  ~~~~0.1962(23)  \\
&  $S_4$  & 1.189(14) & $-0.289$ & 0.963(22) &  ~~~~0.1811(23)  \\
\end{tabular}
\caption{The dimensionless Sommer scale and the coefficients in the fit 
of the dimensionless $\Sigma_g^+$ potential to the Cornell potential in Eq.~\eqref{Cornell-Bicudo} 
for each of the 4 ensembles in Ref.~\cite{Morningstar}
and each of the 4 ensembles in Ref.~\cite{Sharifian:2023idc}.
The value of $r_0\sqrt{\sigma_e}$ is obtained using a smooth interpolation function for all the points 
as described in Section~\ref{sec:scales}.
The coefficient $\kappa_0$ is fixed at the central value from  Eq.~\eqref{Cornell-SW}.
The parameter $\sigma_0/\sigma_e$ is determined from Eq.~\eqref{sigma0/sigma}.
The value of $E'_0/\sqrt{\sigma_e}$ is then determined by 
a 1-parameter  fit of the potential to Eq.~\eqref{Cornell-Bicudo} using the central value of  $\sigma_0/\sigma_e$.
The error bars come only from the statistical errors in the dimensionless potentials $V(r)/\sqrt{\sigma_e}$.
}
\label{tab:Cornellcoeffs}
\end{table}

Step 4:
The transformed dimensionless potential $\tilde{V}_{\Lambda_\eta^\epsilon}^e(r)\, r_0$
corresponding to a dimensionless potential $V_{\Lambda_\eta^\epsilon}^e(r)/\sqrt{\sigma_e}$
from an ensemble $e$ in  Refs.~\cite{Morningstar,Bicudo:2021tsc,Sharifian:2023idc} is
\beq
\tilde{V}_{\Lambda_\eta^\epsilon}^e(r)\,  r_0\ = \left(  V_{\Lambda_\eta^\epsilon}^e(r)\big/\sqrt{\sigma_e} 
- \frac{E_0^\prime}{\sqrt{\sigma_e}} 
+ \frac{r_0E_0}{r_0 \sqrt{\sigma_e}} \right) (r_0 \sqrt{\sigma_e}\,),
\label{V-transform}
\eeq
where the dimensionless constant $r_0E_0$ is given in Eq.~\eqref{Cornell-SW}
and the dimensionless constants $E_0^\prime/\sqrt{\sigma_e}$ and $r_0 \sqrt{\sigma_e}$ 
 for the ensemble $e$ are given in Table~\ref{tab:Cornellcoeffs}.
The rescaled radial variable is $r/r_0 = r\sqrt{\sigma_e}/(r_0 \sqrt{\sigma_e})$.
The transformed  dimensionless potential 
in Eq.~\eqref{V-transform} can be compared directly 
with a corresponding potential in Refs.~\cite{Capitani:2018rox,Schlosser:2021wnr}.
We take the statistical error in the transformed potential to be the product of the statistical error
in $V_{\Lambda_\eta^\epsilon}^e(r)\big/\sqrt{\sigma_e}$ and the central value of $r_0 \sqrt{\sigma_e}$.
The errors from the parameters of the transformation in Eq.~\eqref{V-transform}
can be regarded as systematic errors that are completely correlated within any ensemble $e$.

\begin{figure}[t]
\centerline{ \includegraphics*[width=7cm,clip=true]{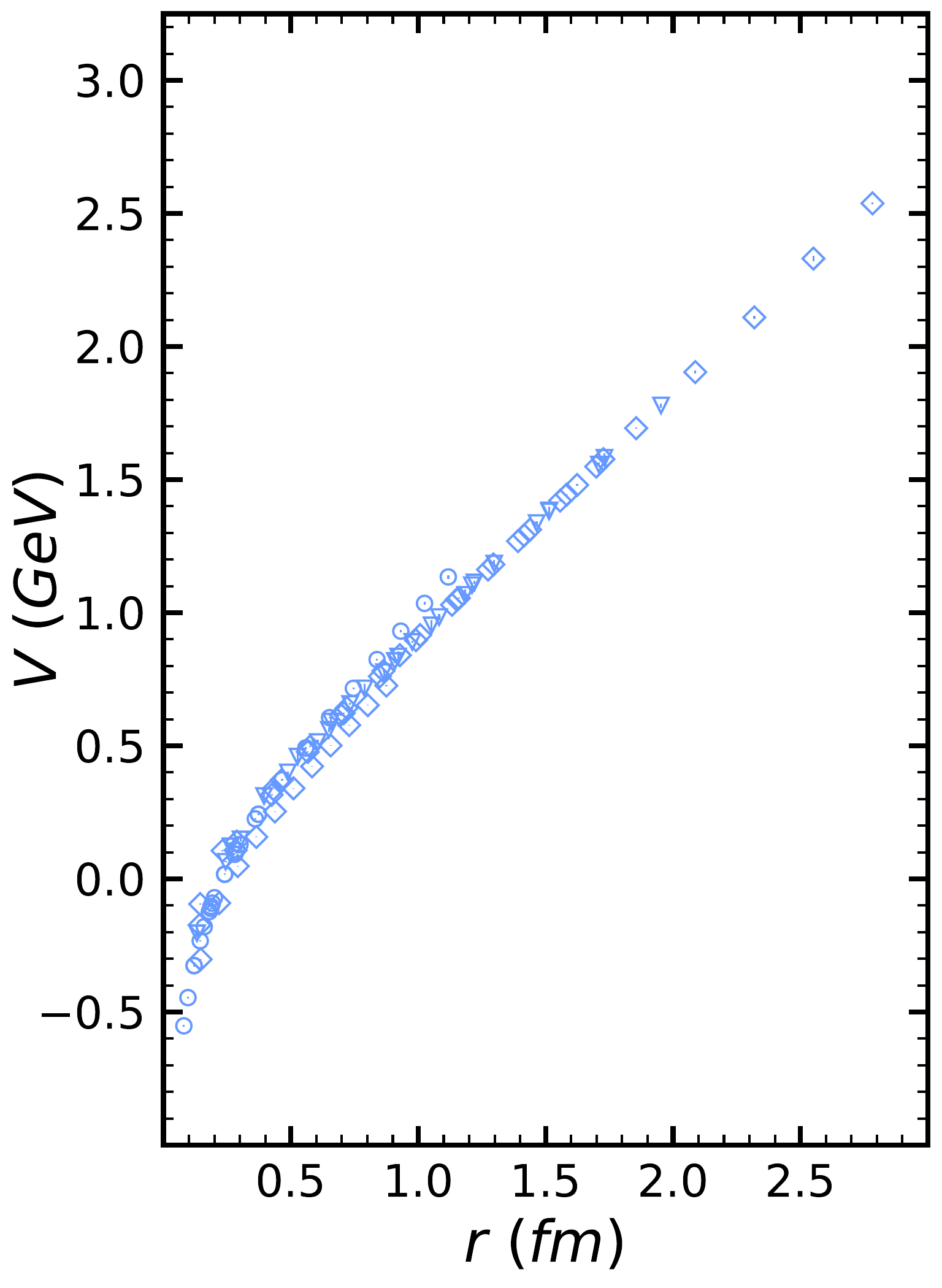}
~ \includegraphics*[width=7cm,clip=true]{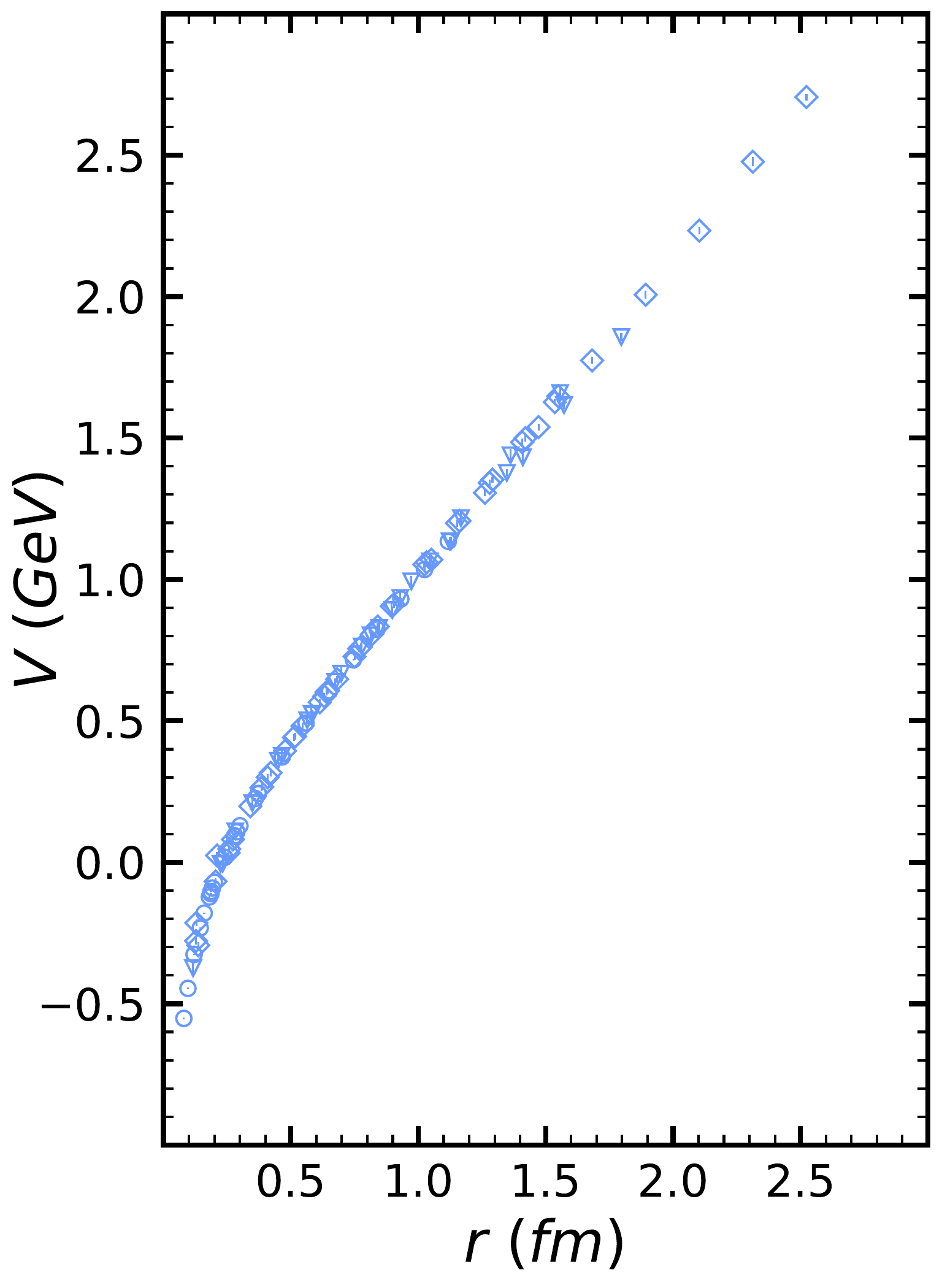} }
\caption{
The ground-state $\Sigma_g^+$ potential for pure $SU(3)$ gauge theory.
The data points are from 
the 4 ensembles in Ref.~\cite{Morningstar} (triangles),
the 5 ensembles in  Refs.~\cite{Schlosser:2021wnr} (circles),
and  the 4 ensembles in Ref.~\cite{Bicudo:2021tsc}  (diamonds).
Left panel:  Physical units  are obtained by setting $r_0 = 0.5$~fm in Ref.~\cite{Schlosser:2021wnr}
and $\sigma_e = 0.18~\mathrm{GeV}^2$ in Refs.~\cite{Morningstar,Bicudo:2021tsc}.
Right panel:  The data points from Refs.~\cite{Morningstar,Bicudo:2021tsc}
have been transformed using Eq.~\eqref{V-transform} with $r_0 = 0.5$~fm.
The error bars on the potentials (which are hardly visible) are the statistical errors only.
}
\label{fig:VSigmag}
\end{figure}

\subsection{B-O Potentials}
\label{sec:BOPotentials}

The $\Sigma_g^+$ potentials from 
the 4 ensembles in Ref.~\cite{Morningstar},
the 5 ensembles in Ref.~\cite{Schlosser:2021wnr},
and the 4  ensembles in Ref.~\cite{Bicudo:2021tsc}
are shown in the left panel of  Fig.~\ref{fig:VSigmag}. 
The data points from the 5 ensembles in  Ref.~\cite{Schlosser:2021wnr}
line up well with each other.
The data points from the 8 ensembles in  Refs.~\cite{Morningstar,Bicudo:2021tsc} do not line up as well 
with those from Ref.~\cite{Schlosser:2021wnr} or with each other.
The same data points after applying the transformation in Eq.~\eqref{V-transform} to  the data  
from the 4 ensembles  in Ref.~\cite{Morningstar} 
 and the 4 ensembles   in Ref.~\cite{Bicudo:2021tsc} 
are shown in the right panel of Fig.~\ref{fig:VSigmag}.
The data points now line up well with those from Ref.~\cite{Schlosser:2021wnr}.
The transformation in Eq.~\eqref{V-transform} was designed so the $\Sigma_g^+$ potentials from 
different  ensembles would line up in the region $\tfrac12 r_0 < r < 2 r_0$.
However they also line up pretty well at smaller $r$ and at larger $r$.

\begin{figure}[t]
\centerline{ \includegraphics*[width=7cm,clip=true]{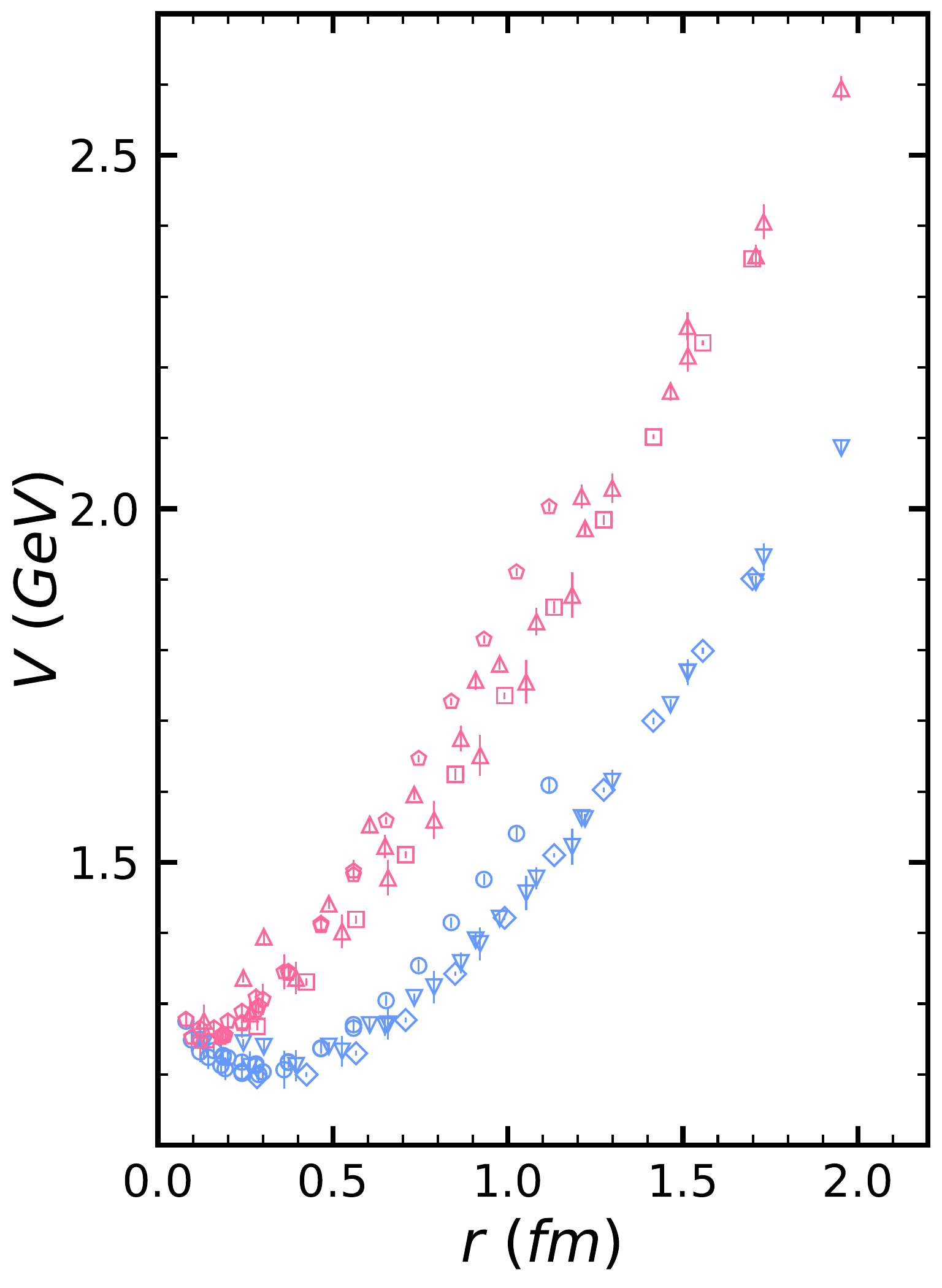} 
~ \includegraphics*[width=7cm,clip=true]{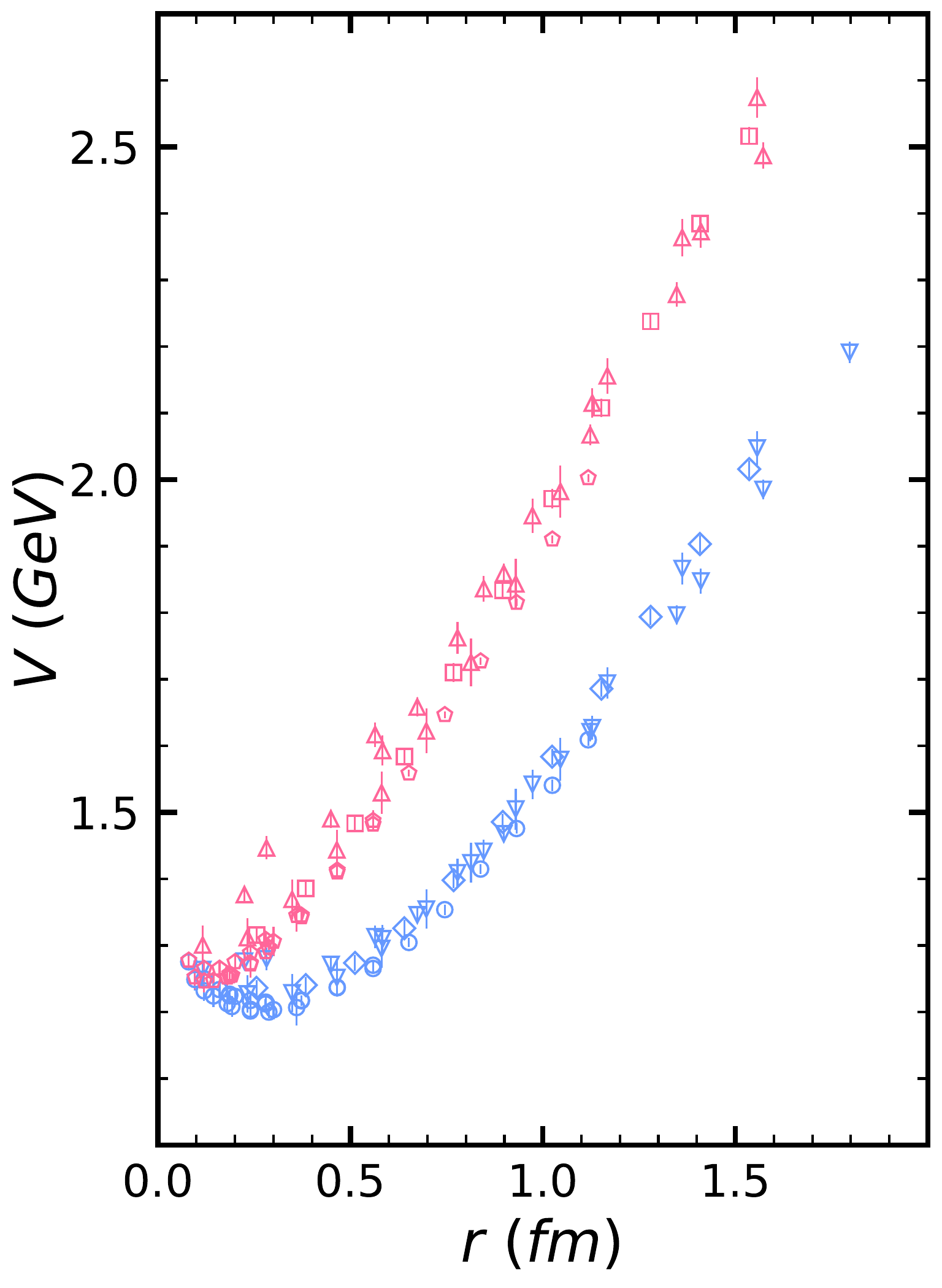} } 
\caption{
The $\Pi_u$ and $\Sigma_u^-$ potentials  for pure $SU(3)$ gauge theory. 
The data points for the $\left(\Pi_u, \Sigma_u^-\right)$ potentials are from 
the 4 ensembles in Ref.~\cite{Morningstar} (down triangles, up triangles),
the 5  ensembles in  Ref.~\cite{Schlosser:2021wnr} (circles, pentagons),
and the $S_4$ ensemble in Ref.~\cite{Sharifian:2023idc} (diamonds, squares).
Left panel:  Physical units  are obtained by setting $r_0 = 0.5$~fm in Ref.~\cite{Schlosser:2021wnr}
and $\sigma _e = 0.18~\mathrm{GeV}^2$ in   Refs.~\cite{Morningstar,Sharifian:2023idc}.
Right panel:  The data points from Refs.~\cite{Morningstar,Sharifian:2023idc}
have been transformed using Eq.~\eqref{V-transform} with $r_0 = 0.5$~fm.
The error bars on the potentials are the statistical errors only.
}
\label{fig:VPiSigmau}
\end{figure}

The $\Pi_u$ and $\Sigma_u^-$ potentials from the 4 ensembles in Ref.~\cite{Morningstar},
the 5 ensembles in Ref.~\cite{Schlosser:2021wnr} and  the $S_4$  ensemble in Ref.~\cite{Sharifian:2023idc}
are shown in the left panel of Fig.~\ref{fig:VPiSigmau}.
The data points for both $\Pi_u$ and $\Sigma_u^-$ from the 5 ensembles in Ref.~\cite{Schlosser:2021wnr} line up well with each other.
The data points for $\Pi_u$ and $\Sigma_u^-$ from the 
4 ensembles in Ref.~\cite{Morningstar} and the $S_4$ ensemble in Ref.~\cite{Sharifian:2023idc}
do not line up well with those  from Ref.~\cite{Schlosser:2021wnr}.
The same data points after applying the transformation in Eq.~\eqref{V-transform} 
are shown in the right panel of Fig.~\ref{fig:VPiSigmau}.
They now line up pretty well with the data points from Ref.~\cite{Schlosser:2021wnr}, 
with the exception of the first few $\Sigma_u^-$ data points from the  ensembles C and D of Ref.~\cite{Morningstar},
which lie above the other $\Sigma_u^-$ data points.
Although the transformation in Eq.~\eqref{V-transform} was designed so the $\Sigma_g^+$ potentials from 
different  ensembles would line up in the region $\tfrac12 r_0 < r < 2 r_0$,
it also makes the data points for both $\Pi_u$ and $\Sigma_u^-$ line up well in all regions of $r$.

\begin{figure}[t]
\centerline{ \includegraphics*[width=7cm,clip=true]{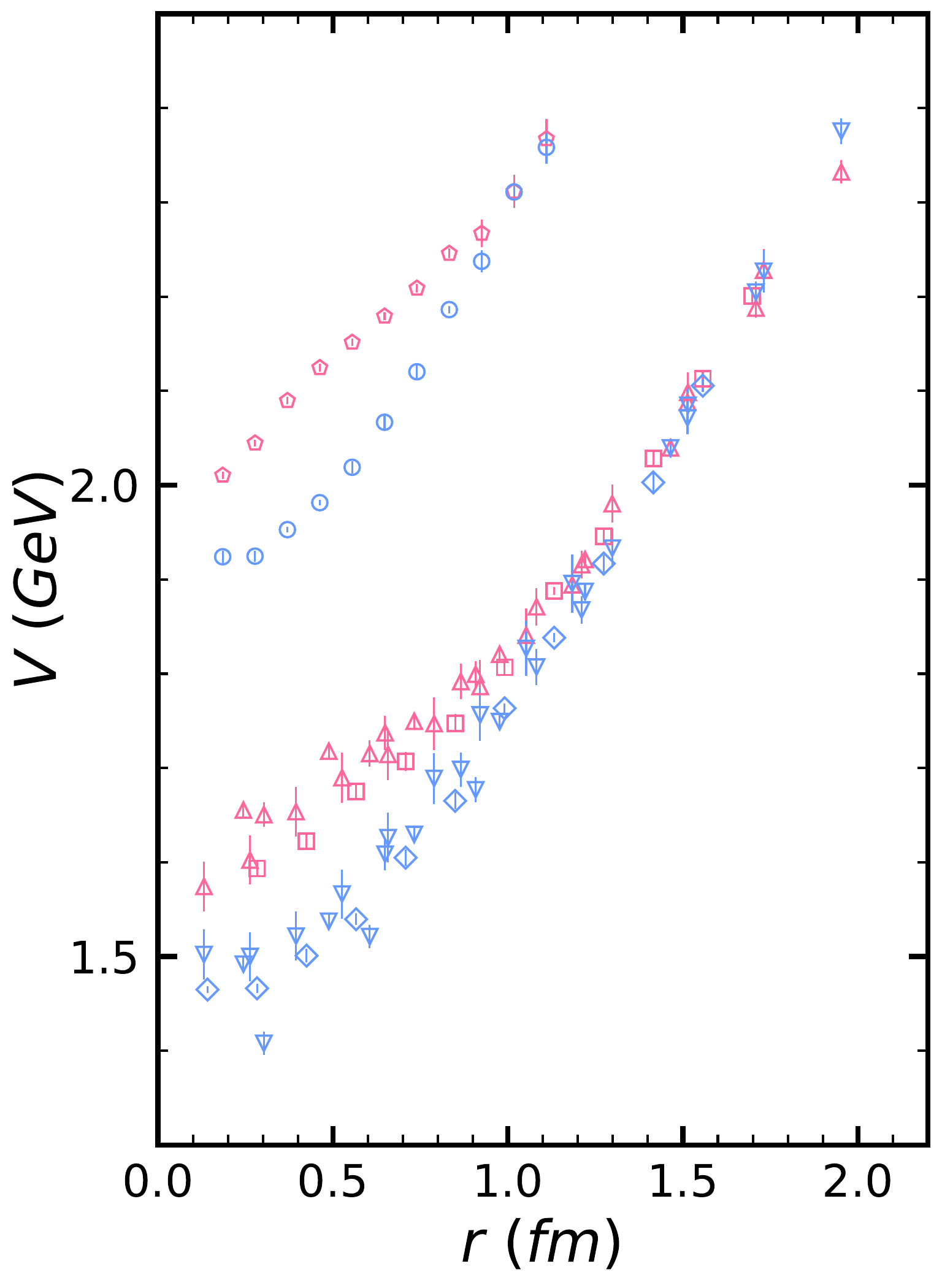}
~ \includegraphics*[width=7cm,clip=true]{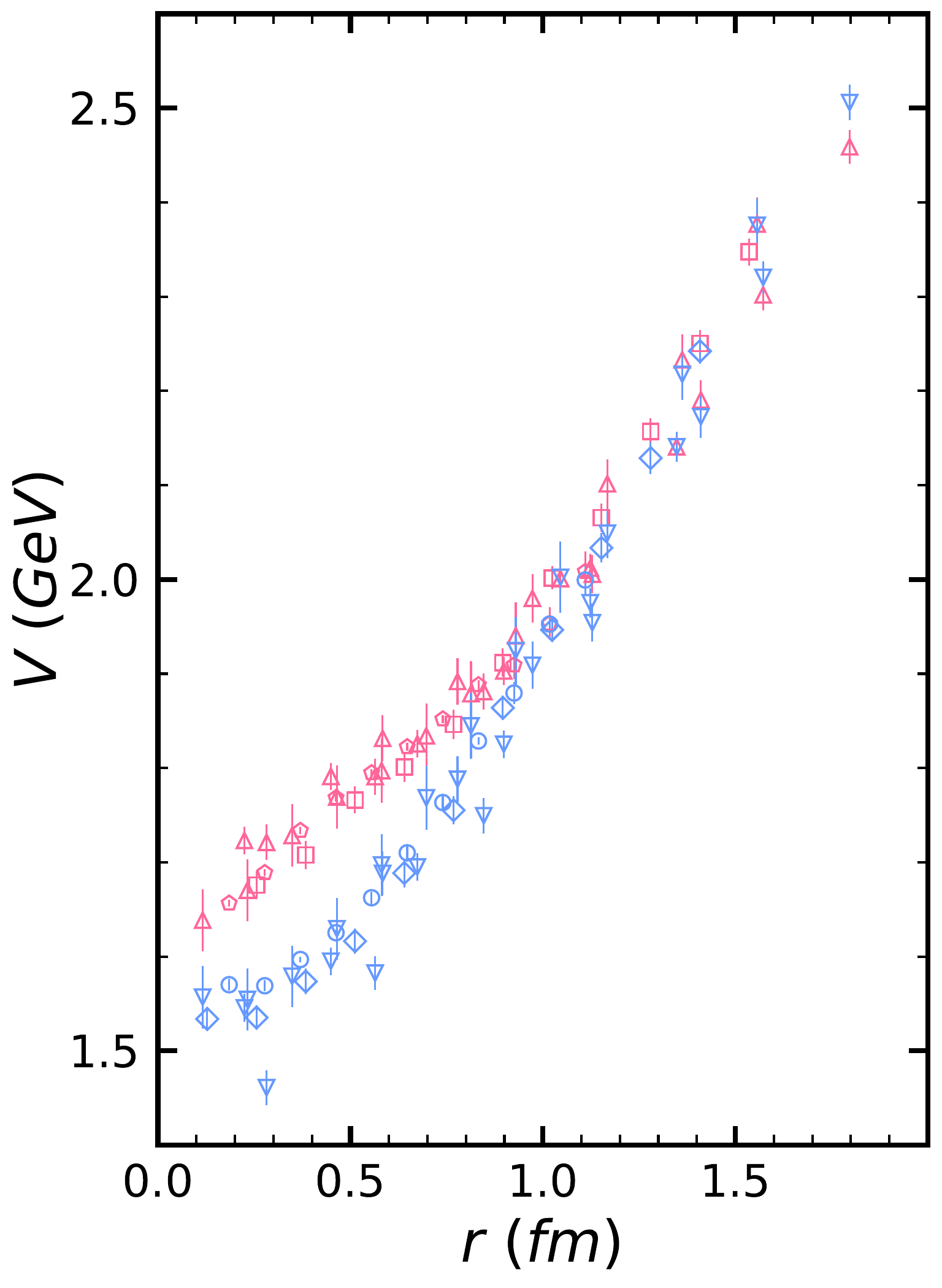} }
\caption{
The  potentials for $\Sigma_g^{+\prime}$ and $\Pi_g$  in pure $SU(3)$ gauge theory. 
The data points for the $(\Sigma_g^{+\prime},\Pi_g)$ potentials are from
 the 4 ensembles in Ref.~\cite{Morningstar} (down triangles, up triangles),
 the HYP2 ensemble in Ref.~\cite{Capitani:2018rox} (circles, pentagons),
and  the 4 ensembles in Refs.~\cite{Bicudo:2021tsc,Sharifian:2023idc} (diamonds, squares).
Left panel:  Physical units are obtained by setting $r_0 = 0.5$~fm in Ref.~\cite{Capitani:2018rox}
and $\sigma _e = 0.18~\mathrm{GeV}^2$ in  Refs.~\cite{Bicudo:2021tsc,Sharifian:2023idc}.
Right panel:  The data points from Refs.~\cite{Bicudo:2021tsc,Sharifian:2023idc, Morningstar}
have been transformed using Eq.~\eqref{V-transform} with $r_0 = 0.5$~fm.
The error bars on the potentials are the statistical errors only.
}
\label{fig:VSigmaPig}
\end{figure}

The $\Sigma_g^{+\prime}$ and $\Pi_g$ potentials from  the 4  ensembles in Ref.~\cite{Morningstar}, 
the HYP2  ensemble in Ref.~\cite{Capitani:2018rox}, 
and  the 4  ensembles in Refs.~\cite{Bicudo:2021tsc,Sharifian:2023idc}
are shown in the left panel of Fig.~\ref{fig:VSigmaPig}.
The same data points after applying the transformation in Eq.~\eqref{V-transform} 
to  the data  in Refs.~\cite{Morningstar,Bicudo:2021tsc,Sharifian:2023idc}
are shown in the right panel of Fig.~\ref{fig:VSigmaPig}.
They now line up pretty well with the $\Sigma_g^{+\prime}$ and $\Pi_g$ data from Ref.~\cite{Capitani:2018rox}
with the exception of  some of the first few $\Sigma_g^{+\prime}$ data points from ensemble D of Ref.~\cite{Morningstar},
which lie below the other $\Sigma_g^{+\prime}$ data points.

\begin{figure}[t]
\centerline{ \includegraphics*[width=7cm,clip=true]{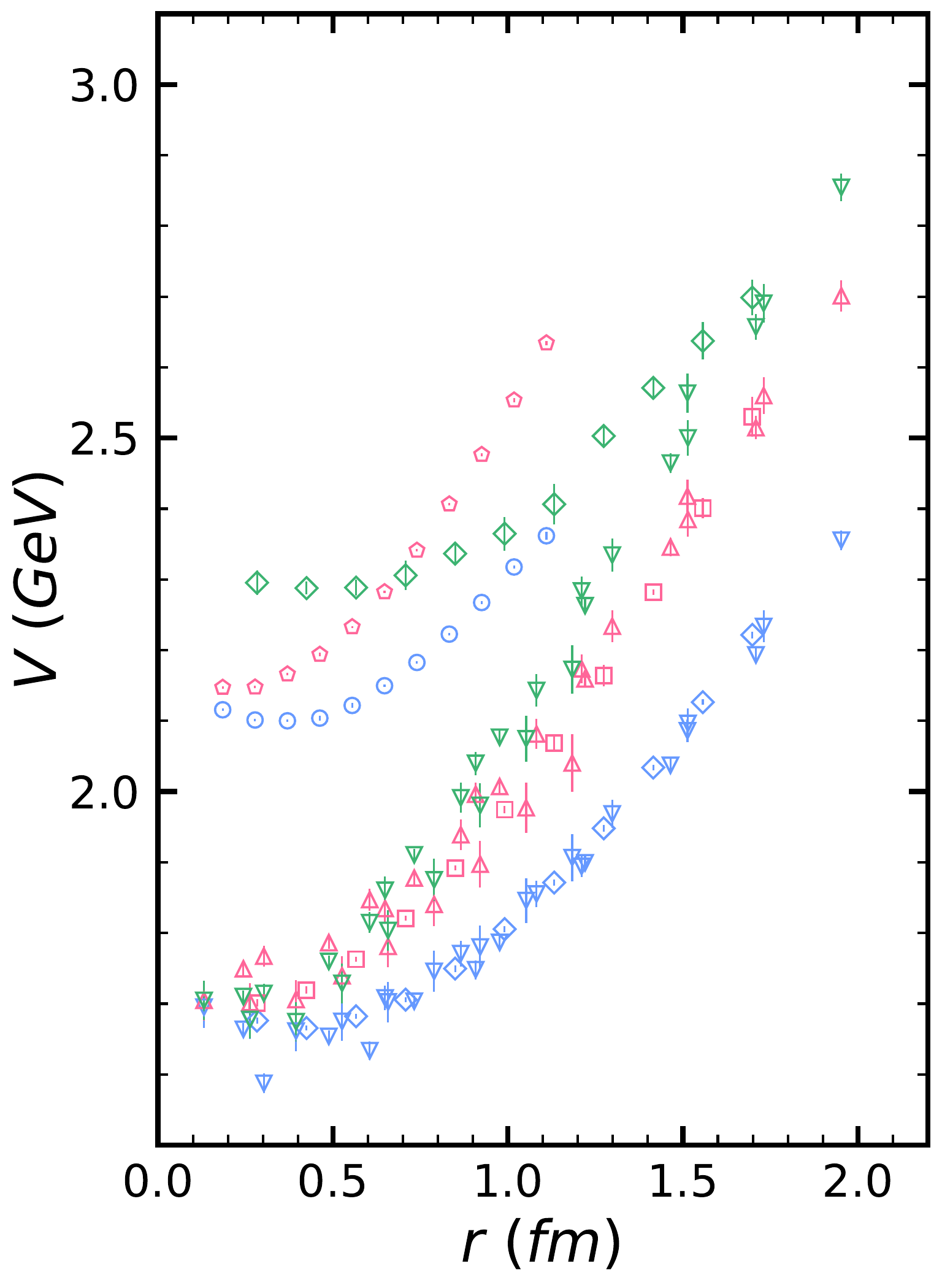}
~ \includegraphics*[width=7cm,clip=true]{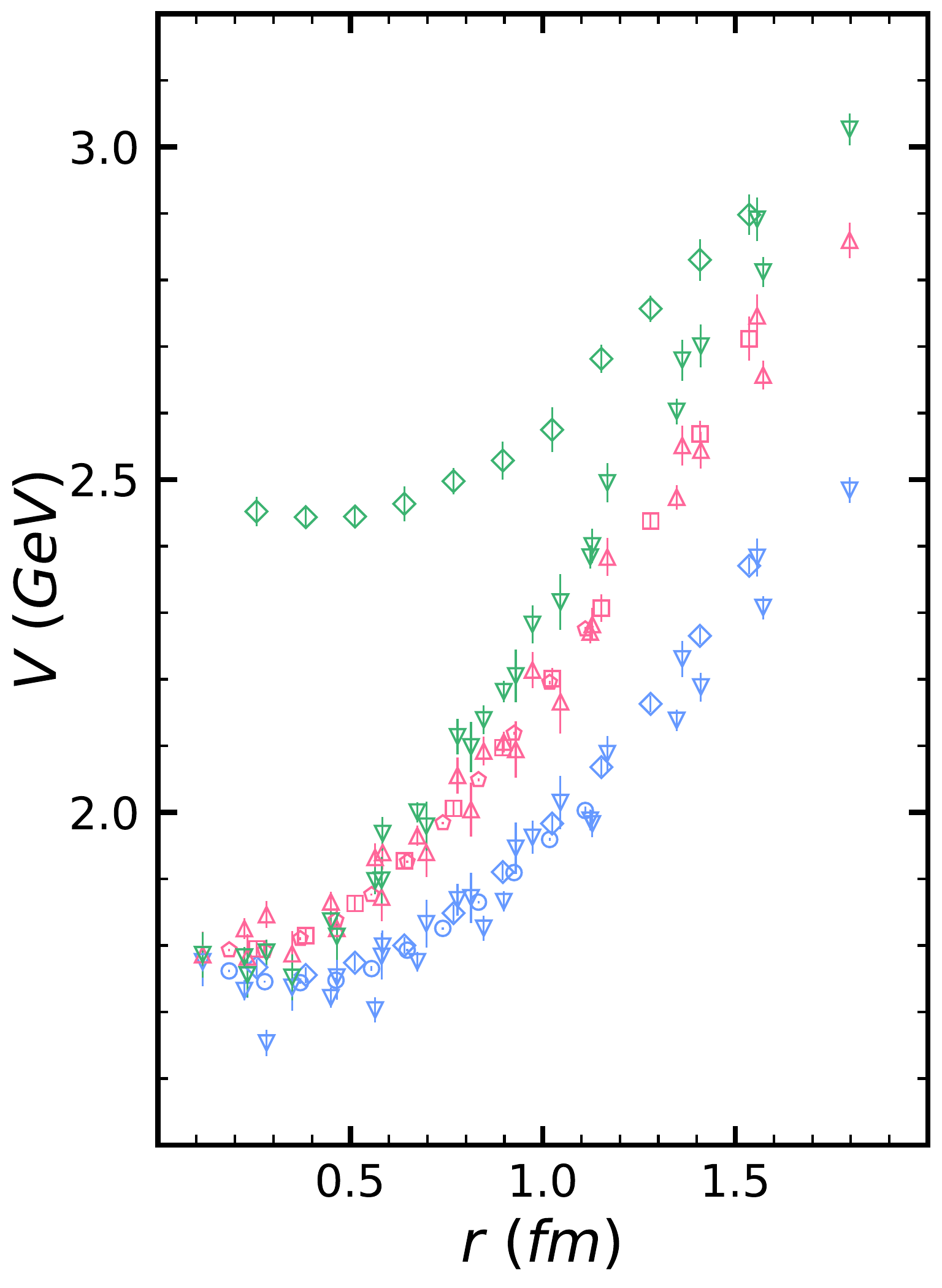} }
\caption{
The  potentials for $\Delta_g$, $\Sigma_g^-$, and $\Pi_g^\prime$ in pure $SU(3)$ gauge theory. 
The data points for the $(\Delta_g,\Sigma_g^-,\Pi_g^\prime)$ potentials are from 
the 4 ensembles in Ref.~\cite{Morningstar} (down triangles, up triangles, higher down triangles),
the HYP2 ensemble in Ref.~\cite{Capitani:2018rox}  (circles, pentagons, none)
and the $S_4$ ensemble in Ref.~\cite{Sharifian:2023idc} (diamonds, squares, higher diamonds).
Left panel:  Physical units are obtained by setting $r_0 = 0.5$~fm  in Ref.~\cite{Capitani:2018rox}
and $\sigma = 0.18~\mathrm{GeV}^2$ in   Refs.~\cite{Morningstar,Sharifian:2023idc}.
Right panel:  The data points from  Refs.~\cite{Morningstar,Sharifian:2023idc}
have been transformed using Eq.~\eqref{V-transform} with $r_0 = 0.5$~fm.
The error bars on the potentials are the statistical errors only.
}
\label{fig:VDeltaPiSigmag}
\end{figure}

The $\Delta_g$, $\Pi_g^\prime$, and $\Sigma_g^-$ potentials from the  4 ensembles in Ref.~\cite{Morningstar},
the HYP2  ensemble in Ref.~\cite{Capitani:2018rox}, and the $S_4$  ensemble in Ref.~\cite{Sharifian:2023idc}  
are shown in the left panel of Fig.~\ref{fig:VDeltaPiSigmag}.
The same data points after applying the transformation in Eq.~\eqref{V-transform} 
to  the  data  from Refs.~\cite{Morningstar,Sharifian:2023idc} 
are shown in the right panel of Fig.~\ref{fig:VDeltaPiSigmag}.
The transformation in Eq.~\eqref{V-transform} lines up the $\Delta_g$ and $\Sigma_g^-$
data points from  Refs.~\cite{Morningstar,Sharifian:2023idc} pretty well
with those from Ref.~\cite{Capitani:2018rox},
with the exception of the first few $\Delta_g$  data points from ensemble D of Ref.~\cite{Morningstar},
which lie below the other $\Delta_g$ data points.

The $\Pi_g^\prime$ potentials in Fig.~\ref{fig:VDeltaPiSigmag} were calculated 
using the 4  ensembles in Ref.~\cite{Morningstar} and the  $S_4$  ensemble in Ref.~\cite{Sharifian:2023idc}.
The $\Pi_g^\prime$ potential was not calculated in Ref.~\cite{Capitani:2018rox}.
At large $r$, the $\Pi_g^\prime$ potentials from  Ref.~\cite{Morningstar} 
and Ref.~\cite{Sharifian:2023idc} are all consistent with approaching the $N=4$ string excitation level.
However they are very different in the small-$r$ region.
The extrapolation of the difference between the $\Pi_g^\prime$ and $\Pi_u$ potentials to $r=0$ can be interpreted 
as the difference between the energies of the $2^{--}$ and $1^{--}$ gluelumps.
For ensemble A of Ref.~\cite{Morningstar}, which is the one with the most data points in the small-$r$ region, 
the extrapolation gives 525(40)~MeV.  
This value is consistent with the gluelump energy difference $\Delta E_{2^{--}} = 523(11)$~MeV for $SU(3)$ gauge theory 
calculated  in Ref.~\cite{Herr:2023xwg}. 
The extrapolation of the difference between the $\Pi_g^\prime$ and $\Pi_u$ potentials in Ref.~\cite{Sharifian:2023idc} 
to $r=0$ gives the much larger value 1119(19)~MeV.
This  value  is inconsistent with the $\Pi_g^\prime$ potential belonging 
to the $2^{--}$ multiplet along with the  $\Delta_g$ and $\Sigma_g^-$ potentials.
The most likely explanation seems to be that a lower-energy state  with B\nobreakdash-O quantum numbers $\Pi_g$
was somehow lost in the calculations of Ref.~\cite{Sharifian:2023idc} in the regions of intermediate  and small $r$.
In the parameterizations of the $\Pi_g^\prime$ potential in Sec.~\ref{sec:SU(3)Potentials},
we will therefore not consider the data from Ref.~\cite{Sharifian:2023idc}.


\section{parameterizations of $\bm{SU(3)}$ Gauge Theory Potentials}
\label{sec:SU(3)Potentials}

In this section, we develop parameterizations of the 8 lowest  B\nobreakdash-O potentials in pure $SU(3)$ gauge theory.

\subsection{General Strategy}
\label{sec:Strategy}

Our parameterizations of the  B\nobreakdash-O potentials  are designed 
to build in the theoretical constraints on their behavior at small $r$ and at large $r$ 
described in Sections~\ref{sec:Vshort} and \ref{sec:Vlong}.
We implement these constraints by using separate parameterizations 
in the regions below and above a matching radius.
For simplicity, the parameterizations at small $r$ are chosen to be linear in the  parameters.
The parameterizations at large $r$ are nonlinear functions of the string tension $\sigma$,
but they are chosen to be linear in the  other parameters.
One  parameter in the small-$r$ expression and one parameter in the large-$r$ 
expression are determined by requiring the potential to be continuous and smooth
at the matching radius.
The  potential is therefore a nonlinear function of $\sigma$ and the matching radius
but a linear function of the remaining adjustable parameters.

We choose the matching radius to have a different value $r_{\Lambda_\eta^\epsilon}$ for each B\nobreakdash-O potential.
The matching radius and the remaining adjustable parameters for the potential
are determined by minimizing the error-weighted  $\chi^2$ for the differences 
between the parameterizations and the potentials calculated using lattice gauge theory.
The ensembles used in the fit and the potentials for each ensemble
are given in Table~\ref{tab:BOpots}.
The potentials for the ensembles HYP2, A, B, C, and D in Refs.~\cite{Capitani:2018rox,Schlosser:2021wnr}
were expressed in terms of the Sommer radius $r_0$.
The potentials for the ensembles A, B, C, and D in Ref.~\cite{Morningstar} 
and the ensembles $W_1$, $W_2$, $W_4$, and $S_4$ in Refs.~\cite{Bicudo:2021tsc,Sharifian:2023idc} 
were expressed in terms of the string tension.
Before these potentials are used in the fits, they are expressed in terms of $r_0$
using the transformation in Eq.~\eqref{V-transform} with the central values of the parameters in Table~\ref{tab:Cornellcoeffs}.
We take the statistical error on the transformed dimensionless potential $\tilde{V}_{\Lambda_\eta^\epsilon}^e(r)\,r_0$ 
to be the  statistical error on the dimensionless potential $V_{\Lambda_\eta^\epsilon}^e(r)\big/\sqrt{\sigma_e}$ 
multiplied by the central value of $r_0 \sqrt{\sigma_e}$.
We  treat the errors from the parameters of the transformation in Eq.~\eqref{V-transform}
as systematic errors that are completely correlated within any ensemble $e$.
We do not attempt to take into account the  correlations between the statistical errors in the potentials and the
systematic errors  from the transformation in Eq.~\eqref{V-transform}.

\begin{table}[t]
\begin{tabular}{ccrrccl}
~References~ &  ~ensemble~ & ~~$n$~ & ~$n_0$ &  ~~~~$r_1$~~~ &  ~~~$r_n$~ ~& ~B-O  potentials \\
\hline
\multirow{4}{*}{\cite{Morningstar}}
& A & 9 & 6 & 0.116 & 1.046 & 
~$\Sigma_g^+$, $\Pi_u$, $\Sigma_u^-$, $\Sigma_g^{+\prime}$, $\Pi_g$, $\Delta_g$, $\Pi_g^\prime$, $\Sigma_g^-$  \\
& B & 6 & 3 & 0.584 & 1.557 & 
~$\Sigma_g^+$, $\Pi_u$, $\Sigma_u^-$, $\Sigma_g^{+\prime}$, $\Pi_g$, $\Delta_g$, $\Pi_g^\prime$, $\Sigma_g^-$  \\
& C & 8 & 3 & 0.225 & 1.797 & 
~$\Sigma_g^+$, $\Pi_u$, $\Sigma_u^-$, $\Sigma_g^{+\prime}$, $\Pi_g$, $\Delta_g$, $\Pi_g^\prime$, $\Sigma_g^-$  \\
& D & 5 & 3 & 0.282 & 1.410 & 
~$\Sigma_g^+$, $\Pi_u$, $\Sigma_u^-$, $\Sigma_g^{+\prime}$, $\Pi_g$, $\Delta_g$, $\Pi_g^\prime$, $\Sigma_g^-$  \\
\hline
\multirow{5}{*}{\cite{Capitani:2018rox}, \cite{Schlosser:2021wnr}} & HYP2 & 11 & 8 & 0.186 & 1.118 & 
~$\Sigma_g^+$, $\Pi_u$, $\Sigma_u^-$, $\Sigma_g^{+\prime}$, $\Pi_g$, $\Delta_g$, $\Sigma_g^-$  \\
& A & 5 & 4 & 0.186 & 0.559 & ~$\Sigma_g^+$, $\Pi_u$, $\Sigma_u^-$  \\
& B & 5 & 2 & 0.120 & 0.360 & ~$\Sigma_g^+$, $\Pi_u$, $\Sigma_u^-$  \\
& C & 5 & 1 & 0.096 & 0.288 & ~$\Sigma_g^+$, $\Pi_u$, $\Sigma_u^-$  \\
& D & 6 & 1 & 0.080 & 0.280 & ~$\Sigma_g^+$, $\Pi_u$, $\Sigma_u^-$ \\
\hline
\multirow{4}{*}{\cite{Bicudo:2021tsc}, \cite{Sharifian:2023idc}} & $W_1$ & 11 & 9 & 0.136 & 0.818 &
~$\Sigma_g^+$, $\Sigma_g^{+\prime}$   \\
& $W_2$ & 12 & 6 & 0.129 & 1.549 & ~$\Sigma_g^+$, $\Sigma_g^{+\prime}$   \\
& $W_4$ & 12 & 3 & 0.210 & 2.523 & ~$\Sigma_g^+$, $\Sigma_g^{+\prime}$    \\
& $S_4$ & 12 & 6 & 0.128 & 1.536 & ~$\Sigma_g^+$, $\Sigma_g^{+\prime}$  \\
& $S_4$ & 11 & 6 & 0.256 & 1.536 & ~$\Pi_u$, $\Sigma_u^-$, $\Pi_g$, $\Delta_g$, $\Pi_g^\prime$, $\Sigma_g^-$ \\
\end{tabular}
\caption{B-O  potentials from $SU(3)$ lattice gauge theory used to determine the parameterizations.
For each ensemble, we give the total number $n$ of data points, 
the number $n_0$ of data points in the range $\tfrac12r_0 < r < 2\,r_0$,
the endpoints $r_1$ and $r_n$ of the range of $r$ in fm (assuming $r_0=0.5$~fm), 
and the B-O  quantum numbers of the potentials.
The endpoints $r_1$ and $r_n$ for the ensembles of 
Refs.~\cite{Morningstar,Bicudo:2021tsc,Sharifian:2023idc}
take into account the scaling factor $r_0 \sqrt{\sigma_e}$ from Table~\ref{tab:Cornellcoeffs}.
}
\label{tab:BOpots}
\end{table}

Our fitting procedure begins with the ground-state $\Sigma_g^+$ potential.
This potential was calculated using all the ensembles listed in Table~\ref{tab:BOpots}.
The calculations for $\Sigma_g^+$ extend out to about 2.8~fm,
which is significantly larger than for any of the other potentials.
This is therefore the best potential for determining the string tension $\sigma$ 
and the energy offset $E_0$ in the expression for the potential at large $r$ in Eq.~\eqref{V-larger}.

Our fitting procedure proceeds with the $\Pi_u$  potential, 
which  is one of the two potentials associated with the $1^{+-}$ gluelump.
Of all the hybrid potentials, the $\Pi_u$  potential is the one 
with the most data points  in the region of $r$ below the minimum of the potential.
It us therefore the best potential for determining the strength $\kappa_8$ of the repulsive color-Coulomb potential
in the expression for a hybrid potential at small $r$ in Eq.~\eqref{VLambda-smallr}.

We simplify the fitting of the higher hybrid potentials 
by fitting instead the differences between the $\Lambda_\eta^\epsilon$ and $\Pi_u$ potentials,
because the color-Coulomb potential $\kappa_8/r$ cancels in the difference between the potentials at small $r$.
We fit the $\Sigma_u^-$ potential which is associated with the $1^{+-}$ gluelump,
the $\Sigma_g^{+\prime}$ and $\Pi_g$ potentials which are associated with the $1^{--}$ gluelump,
and the $\Delta_g$, $\Pi_g^\prime$, and $\Sigma_g^-$ potentials,
which are associated with the $2^{--}$ gluelump.
We  find that there seems to be a narrow avoided crossing between the $\Pi_g$ and $\Pi_g^\prime$ potentials.

\subsection{$\bm{\Sigma_g^+}$ potential}
\label{sec:SigmagpPotential}

We parametrize the $\Sigma_g^+$  potential
by simple analytic expressions below and above a matching radius $r_{\Sigma_g^+}$.
A simple parametrization that is consistent with Eq.~\eqref{VSigmag+-smallr} at small $r$
and with Eq.~\eqref{V-larger}  with string excitation level $N=0$ at large $r$ is
\begin{subequations}
\bqa
V_{\Sigma_g^+}(r) &=&\frac{\kappa_1}{r} + E_{0^{++}} + A_{\Sigma_g^+} \,r^2 
+ B_{\Sigma_g^+} \,r^4+ C_{\Sigma_g^+} \,r^6 \qquad  r < r_{\Sigma_g^+},\\
 &=&  V_0(r)  + E_0 + \frac{D_{\Sigma_g^+}}{r^4}  \qquad\qquad\qquad\qquad\qquad ~~~  r > r_{\Sigma_g^+}.
\eqa
\label{VSigmag}
\end{subequations}
To avoid complex values of $V_0(r)$, we impose the constraint  $r_{\Sigma_g^+} > \sqrt{\pi/(6 \sigma)}$.
We determine the coefficients $C_{\Sigma_g^+}$ and $D_{\Sigma_g^+}$ by requiring 
continuity and smoothness at $r=r_{\Sigma_g^+}$:
\begin{subequations}
\bqa
&& \frac{\kappa_1}{r_{\Sigma_g^+}} + E_{0^{++}}  + A_{\Sigma_g^+} \,r_{\Sigma_g^+}^2 
+ B_{\Sigma_g^+} \,r_{\Sigma_g^+}^4+ C_{\Sigma_g^+} \,r_{\Sigma_g^+}^6 
= V_0(r_{\Sigma_g^+})  + E_0 + \frac{D_{\Sigma_g^+}}{r_{\Sigma_g^+}^4},
\\
&&  - \frac{\kappa_1}{r_{\Sigma_g^+}^2 }  + 2\, A_{\Sigma_g^+} \,r_{\Sigma_g^+} 
+ 4\, B_{\Sigma_g^+} \,r_{\Sigma_g^+}^3 + 6\, C_{\Sigma_g^+} \,r_{\Sigma_g^+}^5
=  \frac{\sigma^2\, r_{\Sigma_g^+}}{V_0(r_{\Sigma_g^+})}  - \frac{4\, D_{\Sigma_g^+}}{r_{\Sigma_g^+}^5}.
\eqa
\label{VSigmag-match}
\end{subequations}
The adjustable parameters are $\sigma$, $E_0$, $\kappa_1$, $E_{0^{++}}$, 
$A_{\Sigma_g^+}$, $B_{\Sigma_g^+}$, and $r_{\Sigma_g^+}$.
Note that the first two coefficients $\kappa_1$ and $E_{0^{++}}$ 
in the small-$r$ expansion are independent adjustable parameters from
the first two coefficients $\sigma$ and $E_0$ in the large-$r$ expansion.
\begin{table}[t]
\resizebox{\textwidth}{!}{%
\begin{tabular}{ccc|ccccccccc}
~$J^{PC}$~ & ~~$\Lambda_\eta^\epsilon$~~ & ~~$N$~~ &
 ~~~~~$r_0^2 \sigma$~~~~~ & ~~~~$r_0 E_0$~~~~ &  ~$\kappa_1$ or $\kappa_8$~ &
~~~$r_0 E_{J^{PC}}$~~~ & ~~~$r_0^3 A_{\Lambda_\eta^\epsilon}$~~~ & ~~~$r_0^5 B_{\Lambda_\eta^\epsilon}$~~~ & ~~~$r_0^7 C_{\Lambda_\eta^\epsilon}$~~~& ~~~$ D_{\Lambda_\eta^\epsilon}/r_0^3$~~~ &  ~$r_{\Lambda_\eta^\epsilon}/r_0$~ \\
\hline
\hline
$0^{++}$ & $\Sigma_g^+$ & 0 & 1.384(2) & $-$0.057(3) & $-$0.240(3) & 0.013(21) & 3.20(13) & $-$4.69(36) & 3.49493    &   0.02874 & 0.713  \\
$1^{+-}$ & $\Pi_u$             & 1 & 1.384       & $-$0.057        & $+$0.037(5) & 2.984(23) & $-$0.010(65) & 0.209(67) & -0.05314    & 0.12717 & 1.298   \\
\end{tabular}}
\caption{
Dimensionless parameters in the $\Sigma_g^+$ and $\Pi_u$ potentials for pure $SU(3)$ gauge theory.
The adjustable parameters in Eqs.~\eqref{VSigmag} and \eqref{VPiu} 
were determined by fitting the data in Figs.~\ref{fig:VSigmag-fit} and \ref{fig:VPiSigmau-fit}.
The errors come only from the statistical errors in the dimensionless potentials. The central values of the parameters $C_{\Lambda_\eta^\epsilon}$ and $D_{\Lambda_\eta^\epsilon}$
determined by matching are also given. The central values of $r_0^2 \sigma$ and $r_0 E_0$ from $\Sigma_g^+$ were used in the fit for $\Pi_u$.
}
\label{tab:SigmagpPiu-params}
\end{table}

\begin{figure}[t]
\centerline{ \includegraphics*[width=12cm,clip=true]{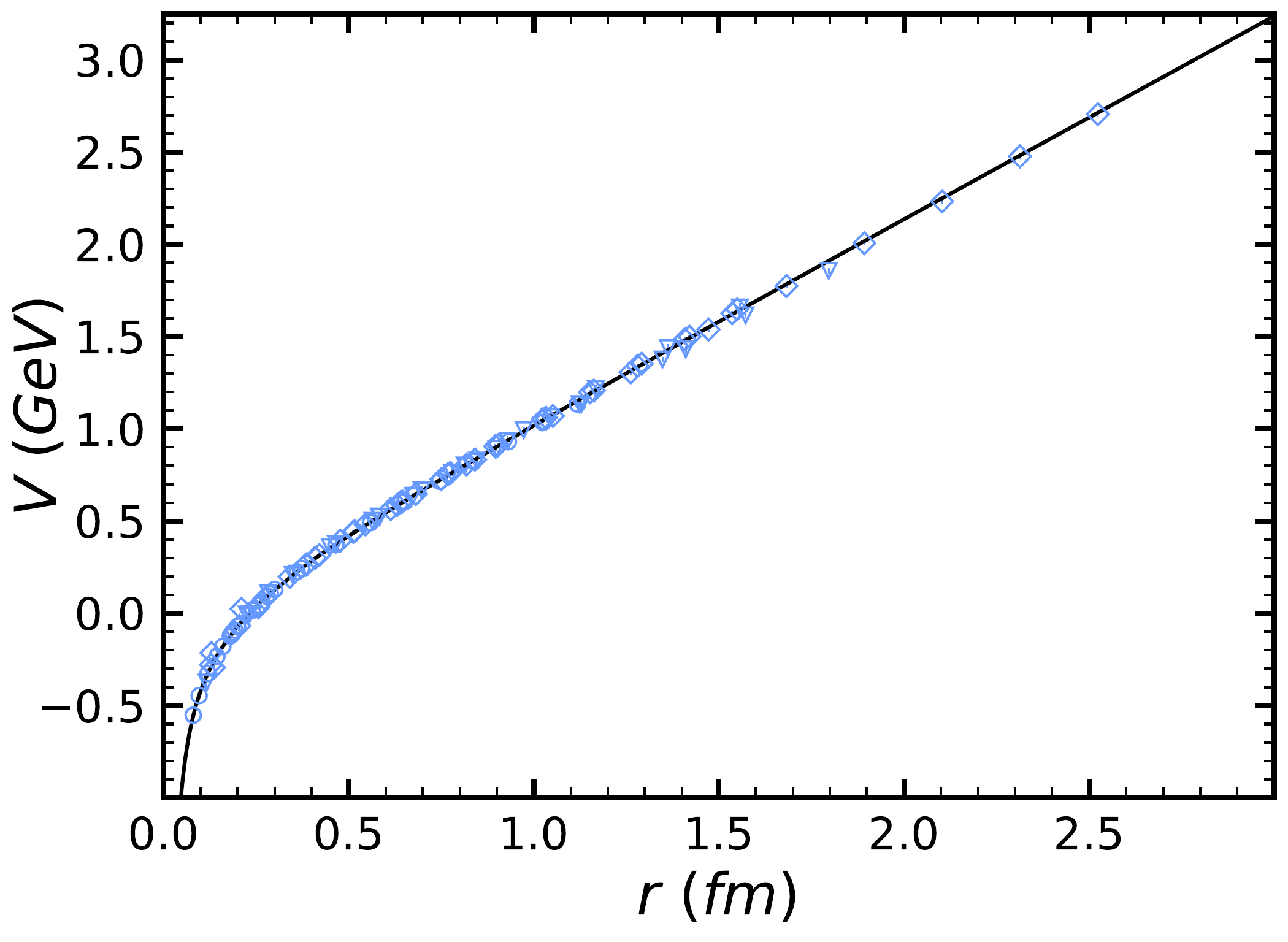} }
\caption{
The ground-state $\Sigma_g^+$ potential for pure $SU(3)$ gauge theory.
The data points are from the 4 ensembles in Ref.~\cite{Morningstar} (triangles), 
the 5 ensembles in Ref.~\cite{Schlosser:2021wnr} (circles),
and the 4 ensembles in Ref.~\cite{Bicudo:2021tsc} (diamonds). 
The data from Refs.~\cite{Morningstar,Bicudo:2021tsc} have been transformed using Eq.~\eqref{V-transform}.
Physical units are obtained by setting $r_0 = 0.5$~fm.
The error bars on the potentials (which are barely visible) are the statistical errors only.
The curve is the potential in Eq.~\eqref{VSigmag} 
with the central values of the parameters in Table~\ref{tab:SigmagpPiu-params}.
}
\label{fig:VSigmag-fit}
\end{figure}

The $\Sigma_g^+$ potential was calculated using all  13  ensembles listed in Table~\ref{tab:BOpots}.
The data points are shown in Fig.~\ref{fig:VSigmag-fit}.
We determine the 6 adjustable parameters in Eq.~\eqref{VSigmag}
by minimizing the error-weighted $\chi^2$ for the 107  data points.
The  fitted dimensionless parameters are given in Table~\ref{tab:SigmagpPiu-params}.
Our fit for the $\Sigma_g^+$ potential is shown in  Fig.~\ref{fig:VSigmag-fit}.

In Ref.~\cite{Capitani:2018rox}, the $\Sigma_g^+$ potential  from the HYP2 ensemble 
before the subtraction of lattice discretization errors was fit to the Cornell potential in Eq.~\eqref{V-Cornell}.
The fit was to
10 points ranging from  0.28 to 1.12~fm. 
The dimensionless string tension was $r_0^2\sigma_0 = 1.387(5)$, which is only 0.2\% larger than our  value in Table~\ref{tab:SigmagpPiu-params}. 
The Coulomb strength was $\kappa_0 = -0.263(2)$, 
which is about 10\%  larger than our value of $\kappa_1$ in Table~\ref{tab:SigmagpPiu-params}.

In Ref.~\cite{Schlosser:2021wnr}, the $\Sigma_g^+$ potential  from the 5 ensembles including HYP2 
before the subtraction of lattice discretization errors were fit to the Cornell potential in Eq.~\eqref{V-Cornell}. 
The fit was to 32 points ranging from 0.080 to 1.12~fm.
The dimensionless string tension was  $r_0^2\sigma_0 = 1.348(5)$,
which is  about 3\% smaller than our value in Table~\ref{tab:SigmagpPiu-params}.
The Coulomb strength was  $\kappa_0 = - 0.289(2)$,
which is  about 20\%  larger than our value of $\kappa_1$ in Table~\ref{tab:SigmagpPiu-params}.
The Cornell potential  in Ref.~\cite{Schlosser:2021wnr} lies consistently below the data in the region below 0.2~fm,
so the data is consistent with a smaller value of $\kappa_1$.

An analytic parameterization of the $\Sigma_g^+$ potential for QCD with two flavors of light quarks 
has been presented in Ref.~\cite{Karbstein:2018mzo}.
The potential was parameterized by different analytic expressions  below  a matching radius $r_1$,
above a matching radius $r_2$, and in the interval between $r_1$ and $r_2$,
with the constraints of continuity and smoothness imposed at the two matching radii.
The potential in the short-distance region $r<r_1$ was the perturbative QCD potential 
expanded to third order in the running coupling constant $\alpha_s(1/r)$.
The potential in the intermediate region $r_1 < r <r_2$ was a quadratic polynomial in $r$.
The potential in the long-distance region $r > r_2$ was the Cornell potential in Eq.~\eqref{V-Cornell}
with coefficients chosen to fit lattice QCD results below the string-breaking region near 1.13~fm.
No parametrization of the potential was given for larger $r$.

\subsection{$\bm{\Pi_u}$ and $\bm{\Sigma_u^-}$ potentials}
\label{sec:1-+Potentials}

\begin{figure}[t]
\centerline{ \includegraphics*[width=12cm,clip=true]{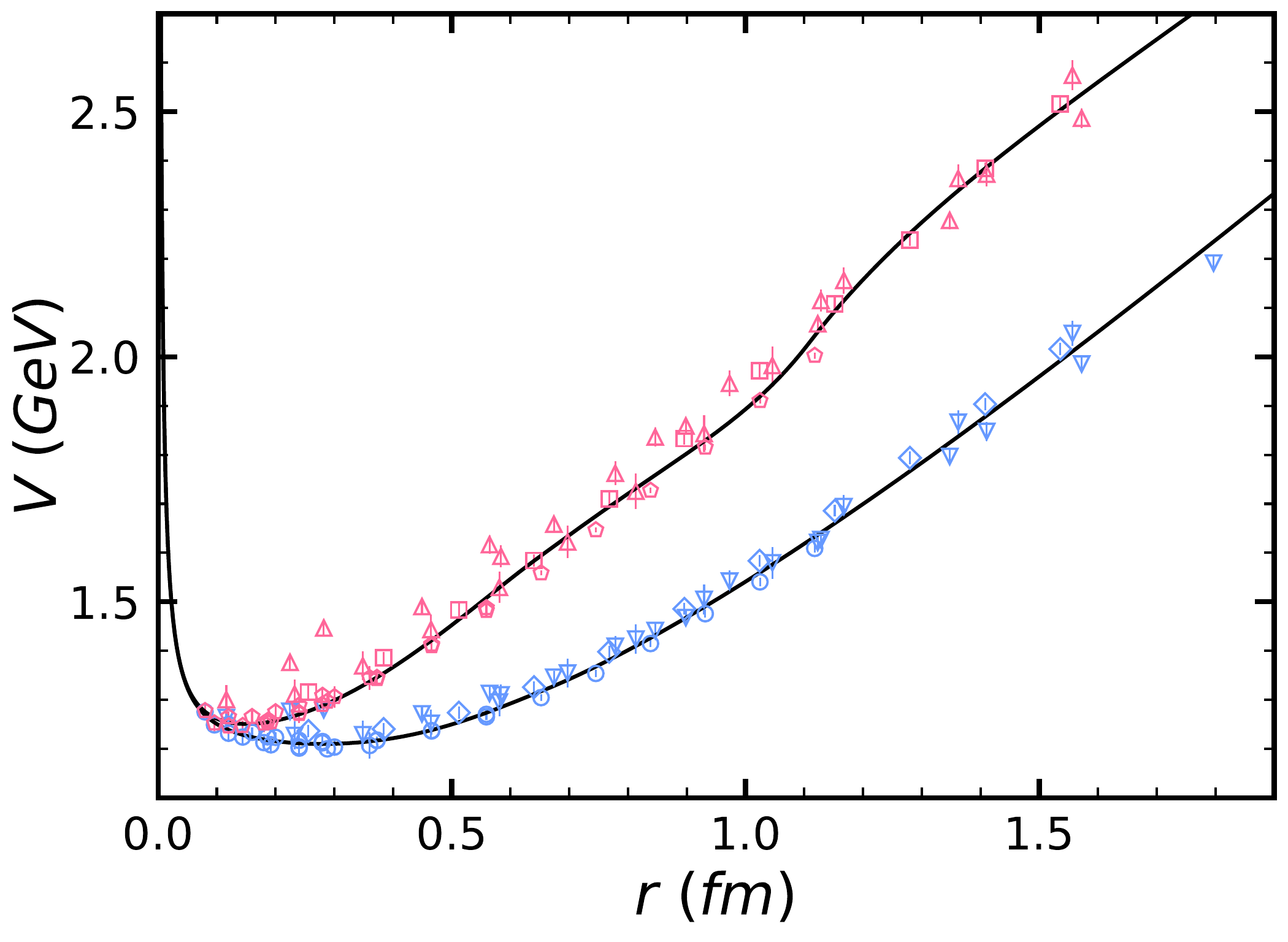} }
\caption{
The $\Pi_u$ and $\Sigma_u^-$ potentials  for pure $SU(3)$ gauge theory. 
The data points for the $(\Pi_u, \Sigma_u^-)$ potentials are from
the 4 ensembles in Ref.~\cite{Morningstar} (down triangles, up triangles),
 the 5  ensembles in Ref.~\cite{Schlosser:2021wnr} (circles, pentagons),
and the $S_4$ ensemble in Ref.~\cite{Sharifian:2023idc} (diamonds, squares).
The data from Refs.~\cite{Morningstar,Sharifian:2023idc} have been transformed using Eq.~\eqref{V-transform}.
Physical units  are obtained by setting $r_0 = 0.5$~fm.
The error bars on the potentials are the statistical errors only.
The lower curve is the $\Pi_u$ potential given by Eq.~\eqref{VPiu} 
with the central values of the parameters in Table~\ref{tab:SigmagpPiu-params}.
The upper curve is the $\Sigma_u^-$ potential given by the sum of the $\Pi_u$ potential 
and the potential difference in Eq.~\eqref{VLambdaeta}
with the central values of the $\Sigma_u^-$ parameters in Table~\ref{tab:Hybrid-params}.
}
\label{fig:VPiSigmau-fit}
\end{figure}

The   $\Pi_u$ and $\Sigma_u^-$   potentials form a multiplet at small $r$ associated with the $1^{+-}$ gluelump.
At large $r$,  the $\Pi_u$ potential approaches the $N=1$ string excitation level.
We parametrize the   $\Pi_u$ potential by simple analytic expressions below and above a matching radius $r_{\Pi_u}$.
A simple parametrization that is consistent with Eq.~\eqref{VLambda-smallr} at small $r$
and with Eq.~\eqref{V-larger}  with $N=1$ at large $r$ is
\begin{subequations}
\bqa
V_{\Pi_u}(r) &=&\frac{\kappa_8}{r} + E_{1^{+-}}  + A_{\Pi_u} \,r^2  
+ B_{\Pi_u} \,r^4 + C_{\Pi_u} \,r^6 \qquad  r < r_{\Pi_u},
\\
 &=&  V_1(r)  + E_0 + \frac{D_{\Pi_u} }{r^4}  \qquad\qquad\qquad\qquad\qquad ~~~  r > r_{\Pi_u}.
\eqa
\label{VPiu}%
\end{subequations}
We determine the coefficients $C_{\Pi_u}$ and $D_{\Pi_u}$ by imposing matching conditions  
at $r=r_{\Pi_u}$ analogous to those in Eqs.~\eqref{VSigmag-match}.
We fix the dimensionless string tension $r_0^2 \sigma$ and the dimensionless energy offset $r_0E_0$
at the central values in Table~\ref{tab:SigmagpPiu-params}.
The adjustable parameters in Eqs.~\eqref{VPiu} are  $\kappa_8$, $E_{1^{+-}}$, $A_{\Pi_u}$, $B_{\Pi_u}$, and $r_{\Pi_u}$.
The $\Pi_u$  potential was calculated using  the 4 ensembles in Ref.~\cite{Morningstar},
the 5 ensembles in Ref.~\cite{Schlosser:2021wnr}, and the $S_4$ ensemble in Ref.~\cite{Sharifian:2023idc}. 
The data points are shown in Fig.~\ref{fig:VPiSigmau-fit}.
We determine the 5 adjustable parameters in Eq.~\eqref{VPiu}
by minimizing the error-weighted $\chi^2$ for the  71 data points.
The fitted dimensionless parameters are given in Table~\ref{tab:SigmagpPiu-params}. 
The ratio of the color-Coulomb coefficients from the $\Sigma_g^+$ and $\Pi_u$ potentials is $\kappa_1/\kappa_8 = -6.5(9)$, which is reasonably close to the tree-level perturbative value $-8$.
Our fit for the $\Pi_u$ potential is shown in Fig.~\ref{fig:VPiSigmau-fit}.

The  other hybrid potentials $\Lambda_\eta^\epsilon$ could be parametrized
in a way analogous to the $\Pi_u$ potential in Eqs.~\eqref{VPiu}.
We choose instead to parametrize the difference between the $\Lambda_\eta^\epsilon$ and $\Pi_u$  potentials
by simple analytic expressions below and above a matching radius $r_{\Lambda_\eta^\epsilon}$.
A simple parametrization of the difference that is consistent with Eq.~\eqref{VLambda-smallr} at small $r$
and with Eq.~\eqref{V-larger}  with string excitation level $N$ at large $r$ is
\begin{subequations}
\bqa
V_{\Lambda_\eta^\epsilon}(r) - V_{\Pi_u}(r) &=&
\Delta E_{J^{PC}}  + \Delta A_{\Lambda_\eta^\epsilon} \,r^2  + \Delta B_{\Lambda_\eta^\epsilon} \,r^4 + \Delta C_{\Lambda_\eta^\epsilon} \,r^6
\qquad  r < r_{\Lambda_\eta^\epsilon},
\\
 &=&  V_N(r) - V_1(r)  +  \frac{\Delta D_{\Lambda_\eta^\epsilon}}{r^4}  
 \qquad\qquad\qquad\qquad\, ~~~  r > r_{\Lambda_\eta^\epsilon},
\eqa
\label{VLambdaeta}%
\end{subequations}
where $\Delta E_{J^{PC}} = E_{J^{PC}} - E_{1^{+-}}$ 
is the energy difference between the $J^{PC}$ and $1^{+-}$ gluelumps.
At small $r$, the color-Coulomb potential $\kappa_8/r$ cancels in the difference between the potentials.
At large $r$, there is a near cancellation between the string potentials $V_N(r)$ and $V_1(r)$.
We determine the coefficients $\Delta C_{\Lambda_\eta^\epsilon}$ and $\Delta D_{\Lambda_\eta^\epsilon}$
 in Eq.~\eqref{VLambdaeta} by imposing matching conditions  
 at $r = r_{\Lambda_\eta^\epsilon}$ analogous to those in Eqs.~\eqref{VSigmag-match}.
We fix the dimensionless string tension  $r_0^2 \sigma$ at the central value in Table~\ref{tab:SigmagpPiu-params}.
We fix the gluelump energy difference $\Delta E_{J^{PC}}$ at the central value in Ref.~\cite{Herr:2023xwg}.
The only  remaining adjustable parameters are 
$\Delta A_{\Lambda_\eta^\epsilon}$, $\Delta B_{\Lambda_\eta^\epsilon}$, 
and $r_{\Lambda_\eta^\epsilon}$.

The $\Sigma_u^-$   potential is the other member besides $\Pi_u$ of the  multiplet at small $r$ associated with the $1^{+-}$ gluelump.
At large $r$,  the $\Sigma_u^-$ potential approaches the $N=3$ string excitation level.
A simple parametrization of the  difference between the  $\Sigma_u^-$  and $\Pi_u$  potentials that is consistent with Eq.~\eqref{VLambda-smallr} 
at small $r$ and with Eq.~\eqref{V-larger}  at large $r$ is
Eq.~\eqref{VLambdaeta} with $\Delta E_{1^{+-}}  = 0$ and $N=3$.
The adjustable parameters are $\Delta A_{\Sigma_u^-}$, $\Delta B_{\Sigma_u^-}$, and $r_{\Sigma_u^-}$.
The  $\Sigma_u^-$ potential was calculated using the 4 ensembles in Ref.~\cite{Morningstar}, 
the  5 ensembles in  Ref.~\cite{Schlosser:2021wnr}
and the $S_4$ ensemble in Ref.~\cite{Sharifian:2023idc}. 
The data points are shown in Fig.~\ref{fig:VPiSigmau-fit}.
We determine the   3 adjustable parameters 
by minimizing the error-weighted $\chi^2$ for the 71 data points.
The  fitted dimensionless parameters are given in Table~\ref{tab:Hybrid-params}.
Our fit for the $\Sigma_u^-$ potential is shown in Fig.~\ref{fig:VPiSigmau-fit}.

In Ref.~\cite{Capitani:2018rox},  parameterizations of the $\Pi_u$ and $\Sigma_u^-$ potentials  from the HYP2 ensemble 
before the subtraction of lattice discretization errors were presented.
The fits were to 11 points for each potential  ranging from about 0.19 to 1.12~fm.
The  parameterization of the $\Pi_u$ potential consisted of the first three terms in 
the small-$r$ expansion in Eq.~\eqref{VLambda-smallr}.
The difference between the $\Sigma_u^-$ and $\Pi_u$ potentials  was expressed as $r^2$ divided by a quadratic 
 polynomial in $r$
with 3 adjustable parameters.
The $\Sigma_u^-$ and $\Pi_u$ potentials both increase like $r^2$ at large $r$, which is inconsistent with stringy constraint in Eq.~\eqref{V-larger}.
The strength of the Coulomb term in the $\Pi_u$ and $\Sigma_u^-$ potentials was $\kappa_8= 0.096(5)$, 
which is larger than our value in Table~\ref{tab:SigmagpPiu-params} by a factor of 2.6.

In Ref.~\cite{Schlosser:2021wnr}, parameterizations of the $\Pi_u$ and $\Sigma_u^-$ potentials  from 5 ensembles 
with lattice discretization errors subtracted were presented.
They used the same parameterizations for the $\Pi_u$ and $\Sigma_u^-$ potentials as in Ref.~\cite{Capitani:2018rox}.
The preferred  fit was to 32 points for each potential ranging from  0.08 to 1.12~fm. The strength of the Coulomb term in the $\Pi_u$ and $\Sigma_u^-$ potentials was $\kappa_8= 0.063(5)$, 
which is larger than our value in Table~\ref{tab:SigmagpPiu-params} by a factor of 1.7. 
An alternative fit  to the 19 points in the  small-$r$ region from 0.08 to 0.28~fm gave $\kappa_8= 0.033(8)$, which is only 11\% smaller than our value in Table~\ref{tab:SigmagpPiu-params}.

Applications of BOEFT to quarkonium hybrid mesons require parameterizations of the $\Pi_u$ and $\Sigma_u^-$ potentials.
In Ref.~\cite{Berwein:2015vca}, the potentials were parameterized by different expressions  
below and above a matching radius $r_1= 0.25$~fm. The potential below  $r_1$ was the sum of a perturbative renormalon-subtracted color-octet potential
and an $r^2$ term with a separate adjustable coefficient  for $\Pi_u$ and $\Sigma_u^-$.
The $\Pi_u$ and $\Sigma_u^-$ potentials above  $r_1$  each had 4 adjustable parameters. 
They included a $1/r$ term and terms that at large $r$ increased linearly with $r$.
The coefficients of $r$ were adjustable parameters that were not constrained to be equal to the string tension.

In the application of BOEFT to quarkonium hybrid mesons in Ref.~\cite{Oncala:2017hop}, 
the $\Sigma_u^-$ potential was parameterized by the Cornell potential in Eq.~\eqref{V-Cornell}. 
The coefficient of $r$  was set equal to the corresponding coefficient in the $\Sigma_g^+$ potential,
which is the string tension. 
The coefficient of $1/r$  was set equal to  the corresponding coefficient in the $\Sigma_g^+$ potential 
multiplied by $-1/8$.  Its value 0.061 was larger than $\kappa_8$ in  Table~\ref{tab:SigmagpPiu-params}
by a factor of 1.65.
The only adjustable parameter in the $\Sigma_g^+$ potential was the constant term.
The parameterization of the $\Pi_u$ potential was more complicated.
The first two terms in the small-$r$ expansion and the first two terms in the large-$r$ expansion 
were the same as for  the $\Sigma_g^+$ potential, but it had 3 additional adjustable parameters.

\subsection{$\bm{\Sigma_g^{+\prime}}$ and $\bm{\Pi_g}$ potentials}
\label{sec:1Sigma_g+'Potential}

\begin{table}[t]
\resizebox{\textwidth}{!}{%
\begin{tabular}{ccc|ccccccc}
~$J^{PC}$~ & ~~~~$\Lambda_\eta^\epsilon$~~~~ &  ~~$N$~~ &
~~~$r_0 \Delta E_{J^{PC}}$~~~ & ~~~$r_0^3 \Delta A_{\Lambda_\eta^\epsilon}$~~~  &  ~~~$r_0^5 \Delta B_{\Lambda_\eta^\epsilon}$~~~&   ~~~~~$r_0^7 \Delta C_{\Lambda_\eta^\epsilon}$~~~~~ &
~~~~~$\Delta D_{\Lambda_\eta^\epsilon}/r_0^3$~~~~~ & ~~$r_{\Lambda_\eta^\epsilon}/r_0$~~  \\
\hline
\hline
\multirow{2}{*}{$1^{+-}$} & $\Pi_u$  & 1 & \multirow{2}{*}{0} & 0 & 0 & 0 & 0  & 1.298\\
& $\Sigma_u^-$ & 3 &  & 0.693(9)& $-$0.200(5) &  0.02068     & -17.0947 & 2.216 & \\
\hline
\multirow{2}{*}{$1^{--}$} 
& $\Sigma_g^{+\prime}$ & 2 & \multirow{2}{*}{0.867} &  0.115(12) & $-$0.023(6) & 0.00079     & 0.75235& 2.223\\
& $\Pi_g$ & 2 &  & 1.122(22) & $-$0.739(26) & 0.13543     &  0.58133 & 1.438\\
\hline
\multirow{3}{*}{$2^{--}$}  
& $\Delta_g$ & 2 & \multirow{3}{*}{1.325} & $-$0.020(5) & $-$0.021(3) & 0.00224     & 1.22304 & 2.158\\
& $\Pi_g^\prime$ & 4 &  & 0.137(19) & 0.008(10)  & $-$0.00187  & $-$10.8934 & 2.290  \\
& $\Sigma_g^-$ & 4 &  & 0.272(4) & $-$0.083(2)  & 0.00825     & $-$20.8911 & 2.401 \\
\hline\hline
$1^{--}$ & $\Pi_{g1}$ & 4 & 0.867 & 1.191(11) & $-$0.606(12) & 0.09864     & $-$9.03104 & 1.813\\
$2^{--}$ & $\Pi_{g2}$ & 2 & 1.325 & 0.061(10) & $-$0.057 & 0.00628     & 1.36802 & 2.006 \\
\hline
\end{tabular}}
\caption{
Dimensionless parameters in the differences between the $\Sigma_u^-$, $\Sigma_g^{+\prime}$, $\Pi_g$, 
$\Delta_g$, $\Pi_g^\prime$, and $\Sigma_g^-$ potentials and the $\Pi_u$ potential for pure $SU(3)$ gauge theory.
The dimensionless gluelump energy differences $r_0 \Delta E_{J^{PC}}$ are the central values from Ref.~\cite{Herr:2023xwg}.
The adjustable parameters in Eqs.~\eqref{VLambdaeta} for the $1^{+-}$,  $1^{--}$, and $2^{--}$  multirows
were determined by fitting the data in Figs.~\ref{fig:VPiSigmau-fit}, \ref{fig:VSigmaPig-fit}, and \ref{fig:VDeltaPiSigmag-fit}
to potentials of the form in Eq.~\eqref{VLambdaeta}.
The adjustable parameters in Eqs.~\eqref{VLambdaeta} for the  $\Pi_{g1}$ and $\Pi_{g2}$ rows
were determined by fitting the data for the $\Pi_g$ and $\Pi_g^\prime$ potentials
in Figs.~\ref{fig:VSigmaPig-fit} and \ref{fig:VDeltaPiSigmag-fit} to the avoided-crossing potentials in Eqs.~\eqref{VPiggp}. The central values of the parameters $\Delta C_{\Lambda_\eta^\epsilon}$ and $\Delta D_{\Lambda_\eta^\epsilon}$ determined by matching are also given.
}
\label{tab:Hybrid-params}
\end{table}

The  $\Sigma_g^{+\prime}$ and $\Pi_g$  potentials form a multiplet at small $r$
associated with the $1^{--}$ gluelump.
At large $r$,  both potentials approach the $N=2$ string excitation level.
In the parameterizations in Eq.~\eqref{VLambdaeta} of the difference between these potentials and the $\Pi_u$ potential,
we fix the dimensionless string tension $r_0^2 \sigma$ at  the central value in Table~\ref{tab:SigmagpPiu-params}
and the dimensionless gluelump energy difference 
at the central value  $r_0 \Delta E_{1^{--}}=0.867$ calculated in Ref.~\cite{Herr:2023xwg}.

\begin{figure}[t]
\centerline{ \includegraphics*[width=10cm,clip=true]{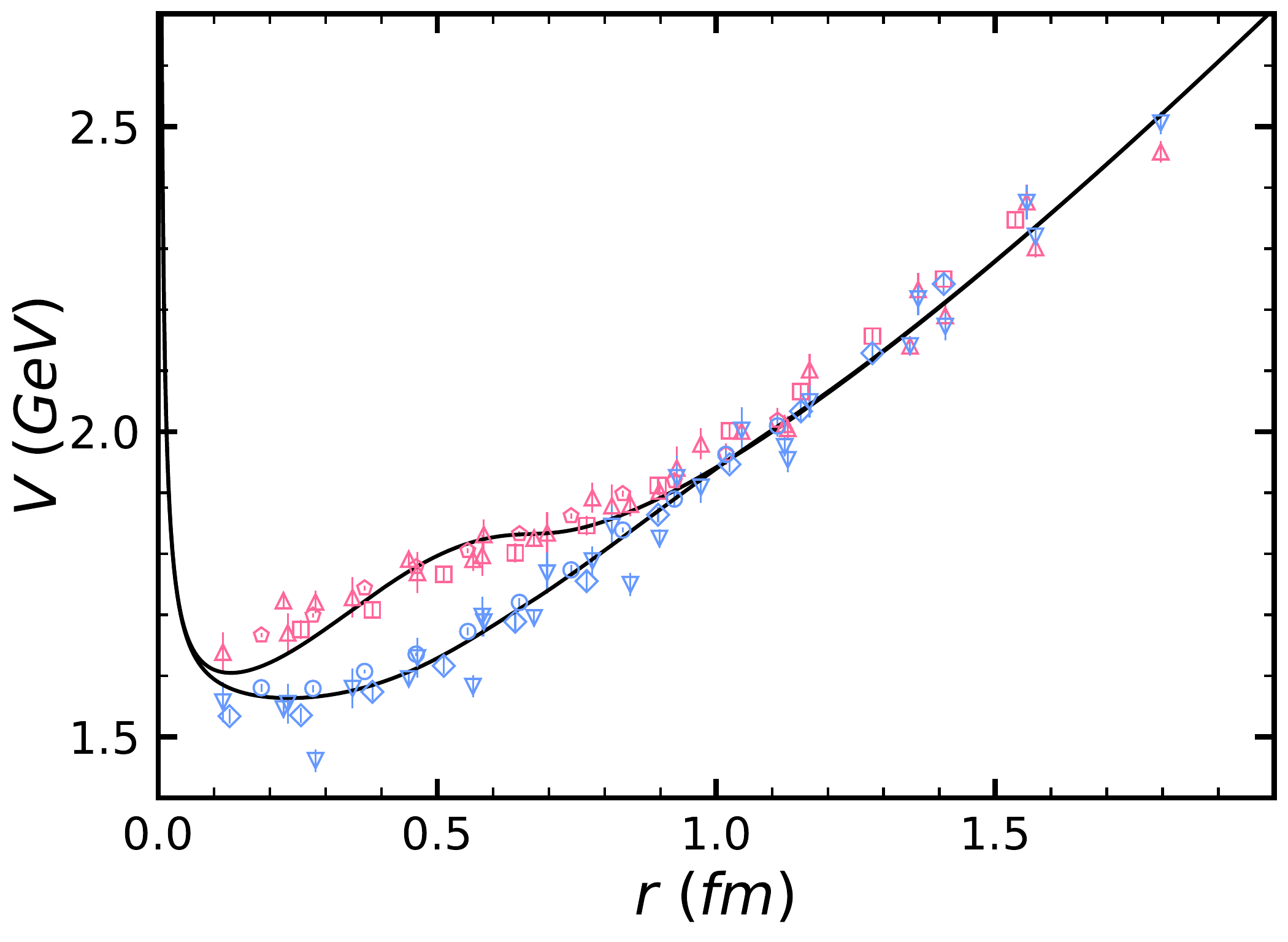} }
\caption{
The  potentials for $\Sigma_g^{+\prime}$ and $\Pi_g$  in pure $SU(3)$ gauge theory. 
The data points for the  $(\Sigma_g^{+\prime},\Pi_g)$ potentials are from 
the 4 ensembles in Ref.~\cite{Morningstar} (down triangles, up triangles),
the HYP2 ensemble in Ref.~\cite{Capitani:2018rox}  (circles, pentagons),
and  the 4 ensembles in Refs.~\cite{Bicudo:2021tsc,Sharifian:2023idc} (diamonds, squares).
The data from Refs.~\cite{Morningstar,Bicudo:2021tsc,Sharifian:2023idc} have been transformed using Eq.~\eqref{V-transform}.
Physical units are obtained by setting $r_0 = 0.5$~fm.
The error bars on the potentials are the statistical errors only.
The lower and upper curves are the $\Sigma_g^{+\prime}$ and $\Pi_g$ potentials given by the sums
 of the $\Pi_u$ potential in Fig.~\ref{fig:VPiSigmau-fit} and the potential differences in Eq.~\eqref{VLambdaeta} 
with the parameters for $\Sigma_g^{+\prime}$ and $\Pi_g$ in Table~\ref{tab:Hybrid-params}.
}
\label{fig:VSigmaPig-fit}
\end{figure}

In the parametrization in Eq.~\eqref{VLambdaeta} of the difference between the  $\Sigma_g^{+\prime}$ and $\Pi_u$ potentials,
the adjustable parameters are $\Delta A_{\Sigma_g^{+\prime}}$, $\Delta B_{\Sigma_g^{+\prime}}$, and $r_{\Sigma_g^{+\prime}}$.
The $\Sigma_g^{+\prime}$ potential was calculated using the $4$ ensembles in Ref.~\cite{Morningstar}, 
the HYP2 ensemble in Ref.~\cite{Capitani:2018rox}, 
and the  $S4$ ensemble in Ref.~\cite{Sharifian:2023idc}.  
The data points are shown in Fig.~\ref{fig:VSigmaPig-fit}.
We determine the 3 adjustable parameters 
by minimizing the error-weighted $\chi^2$ for the 
50 data points.
The  fitted dimensionless parameters are  given in Table~\ref{tab:Hybrid-params}.
Our fit for the $\Sigma_g^{+\prime}$  potential is shown in Fig.~\ref{fig:VSigmaPig-fit}.

In the parametrization in Eq.~\eqref{VLambdaeta} of the difference between the $\Pi_g$ and $\Pi_u$ potentials,
 the adjustable parameters are $\Delta A_{\Pi_g}$, $\Delta B_{\Pi_g}$, and $r_{\Pi_g}$.
The $\Pi_g$ potential was calculated using the 4 ensembles in Ref.~\cite{Morningstar}, the HYP2 ensemble in Ref.~\cite{Capitani:2018rox}, 
and the  $S_4$ ensemble in Ref.~\cite{Sharifian:2023idc}.
The data points are shown in Fig.~\ref{fig:VSigmaPig-fit}.
We determine the 3 adjustable parameters 
by minimizing the error-weighted $\chi^2$ for the 50 data points.
The  fitted dimensionless parameters are  given in Table~\ref{tab:Hybrid-params}.
Our fit for the $\Pi_g$  potential is shown in Fig.~\ref{fig:VSigmaPig-fit}.

Parameterizations of the $\Sigma_g^{+\prime}$ and $\Pi_g$ potentials have been presented previously in Ref.~\cite{Pineda:2019mhw}. They presented alternative parametrizations that increased at large $r$ as $r^2$ or as $r$.

\subsection{$\bm{\Delta_g}$, $\bm{\Pi_g^\prime}$, and $\bm{\Sigma_g^-}$ potentials}
\label{sec:1--Potentials}

The  $\Delta_g$, $\Pi_g^\prime$,  and $\Sigma_g^-$  potentials form a multiplet at small $r$
associated with the $2^{--}$ gluelump.
At large $r$, these potential approach the $N=2$, 4, and 4  string excitation levels, respectively.
In the parameterizations of the potentials in Eq.~\eqref{VLambdaeta},
we fix the dimensionless string tension $r_0^2 \sigma$ at the central value in Table~\ref{tab:SigmagpPiu-params}
and the dimensionless gluelump energy difference at the central value 
$r_0 \Delta E_{1^{--}}=1.325$ calculated in Ref.~\cite{Herr:2023xwg}.

\begin{figure}[t]
\centerline{ \includegraphics*[width=10cm,clip=true]{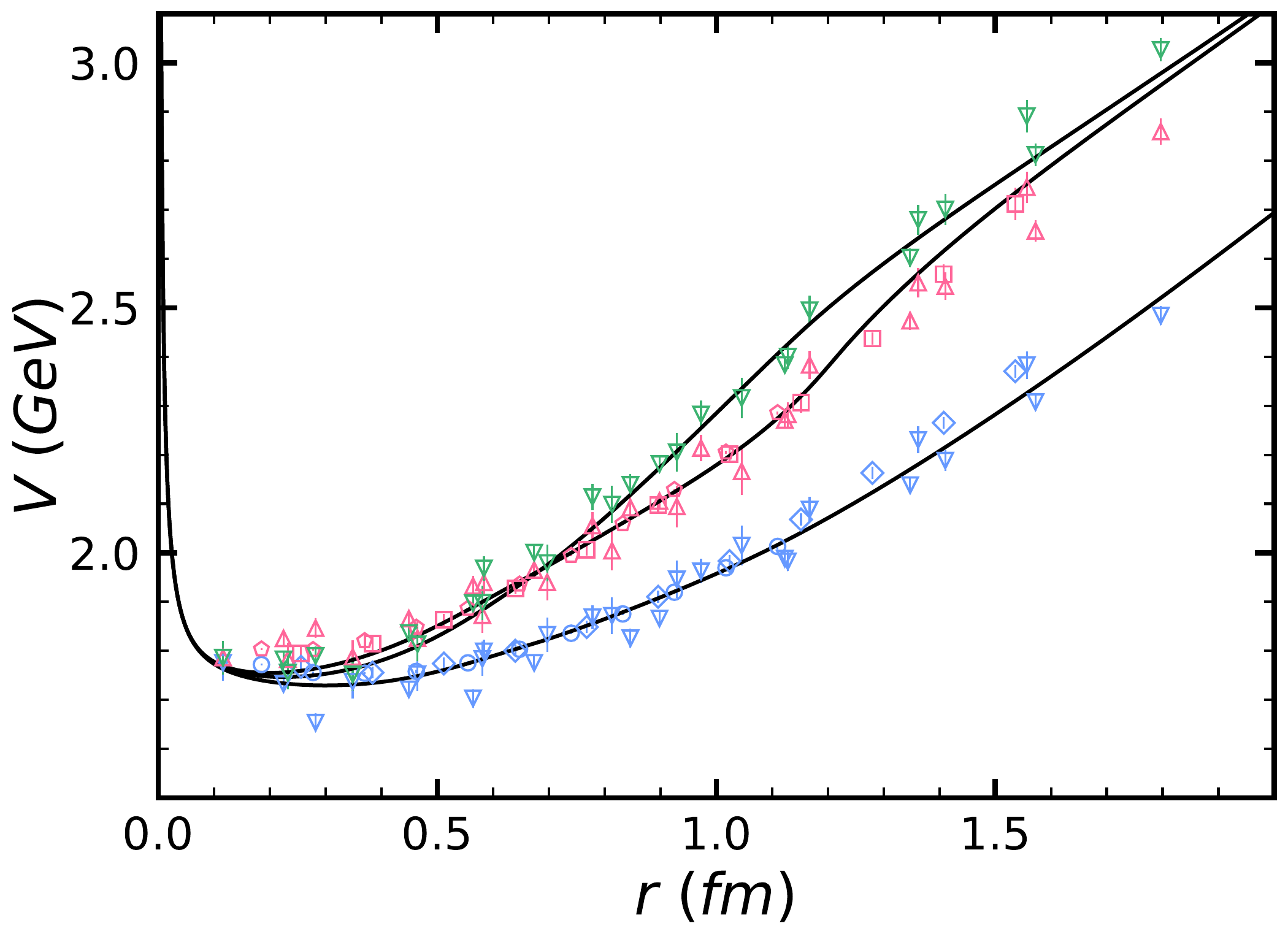} }
\caption{
The  potentials for $\Delta_g$, $\Sigma_g^-$, and $\Pi_g^\prime$ in pure $SU(3)$ gauge theory. 
The data points for the $(\Delta_g,\Sigma_g^-,\Pi_g^\prime)$ potentials are from 
the 4 ensembles in Ref.~\cite{Morningstar} (down triangles, up triangles, higher down triangles),
the HYP2 ensemble in Ref.~\cite{Capitani:2018rox} (circles, pentagons, none),
and  the $S_4$  ensemble in Ref.~\cite{Sharifian:2023idc}  (diamonds, squares, none)
The data from Refs.~\cite{Morningstar,Sharifian:2023idc} have been transformed using Eq.~\eqref{V-transform}.
Physical units  are obtained by setting $r_0 = 0.5$~fm.
The error bars on the potentials are the statistical errors only.
The lower, middle,  and upper curves at large $r$ are the $\Delta_g$,  $\Sigma_g^-$, and $\Pi_g^\prime$ potentials 
given by the sums of the $\Pi_u$ potential in Fig.~\ref{fig:VPiSigmau-fit} and
the potential differences in Eq.~\eqref{VLambdaeta} 
with the parameters for $\Delta_g$,  $\Sigma_g^-$, and $\Pi_g^\prime$ in Table~\ref{tab:Hybrid-params}.
}
\label{fig:VDeltaPiSigmag-fit}
\end{figure}

In the parametrization  in Eq.~\eqref{VLambdaeta} of the difference between the  $\Delta_g$ and $\Pi_u$ potentials, 
the adjustable parameters are $\Delta A_{\Delta_g}$, $\Delta B_{\Delta_g}$, and $r_{\Delta_g}$.
The $\Delta_g$ potential was calculated 
using the HYP2 ensemble in Ref.~\cite{Capitani:2018rox}, 
the 4 ensembles in Ref.~\cite{Morningstar}, and the $S_4$ ensemble in Ref.~\cite{Sharifian:2023idc}.
The data points are shown in  Fig.~\ref{fig:VDeltaPiSigmag-fit}.
We determine the 3 adjustable parameters 
by minimizing the error-weighted $\chi^2$ for the 50 data points.
The  fitted parameters are  given in Table~\ref{tab:Hybrid-params}.
Our fit for the $\Delta_g$  potential is shown in Fig.~\ref{fig:VDeltaPiSigmag-fit}.

In the parametrization in Eq.~\eqref{VLambdaeta} of the difference between the  $\Pi_g^\prime$ and $\Pi_u$ potentials, 
the adjustable parameters are  $\Delta A_{\Pi_g^\prime}$, $\Delta B_{\Pi_g^\prime}$, and $r_{\Pi_g^\prime}$.
The $\Pi_g^\prime$ potential was calculated using the 4 ensembles in Ref.~\cite{Morningstar}.
The data points from Ref.~\cite{Morningstar} are shown in  Fig.~\ref{fig:VDeltaPiSigmag-fit}.
We determine the 3 adjustable parameters 
by minimizing the error-weighted $\chi^2$ for the 28 data points.
The  fitted dimensionless parameters are  given in Table~\ref{tab:Hybrid-params}.
Our fit for the $\Pi_g^\prime$  potential is shown in Fig.~\ref{fig:VDeltaPiSigmag-fit}.

In the parametrization in Eq.~\eqref{VLambdaeta} of the difference between the $\Sigma_g^-$  and $\Pi_u$ potentials, 
the adjustable parameters are $\Delta A_{\Sigma_g^-}$, $\Delta B_{\Sigma_g^-}$, and $r_{\Sigma_g^-}$.
The $\Sigma_g^-$ potential was calculated using the 4 ensembles in Ref.~\cite{Morningstar},
the HYP2 ensemble in Ref.~\cite{Capitani:2018rox}, 
and the $S_4$ ensemble in Ref.~\cite{Sharifian:2023idc}.
The data points are shown in  Fig.~\ref{fig:VDeltaPiSigmag-fit}.
We determine the  3 adjustable parameters 
by minimizing the error-weighted $\chi^2$ for the 50  data points.
The  fitted parameters are  given in Table~\ref{tab:Hybrid-params}.
Our fit for the $\Sigma_g^-$ potential is shown in Fig.~\ref{fig:VDeltaPiSigmag-fit}.
Note that the $\Pi_g^\prime$ and $\Sigma_g^-$ potentials in  Fig.~\ref{fig:VDeltaPiSigmag-fit} cross near 0.7~fm.

\subsection{$\bm{\Pi_g}$ and $\bm{\Pi_g}^\prime$  potentials}
\label{sec:PigPotentials}

\begin{figure}[t]
\centerline{ \includegraphics*[width=12cm,clip=true]{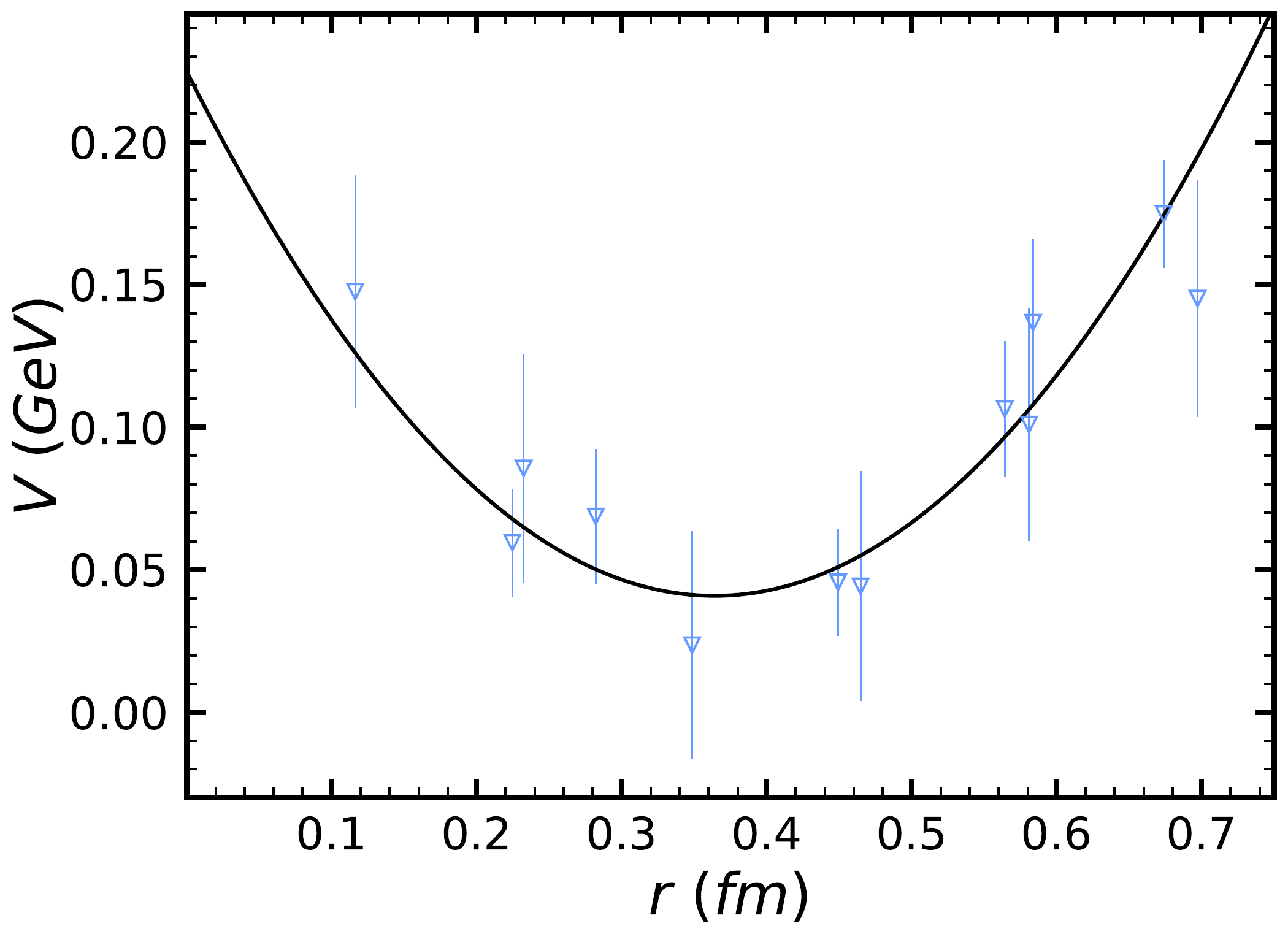}}
\caption{
The difference between the adiabatic $\Pi_g^\prime$ and $\Pi_g$ potentials as a function of $r$.
The data points are the differences between the potentials 
from the 4 ensembles  in Ref.~\cite{Morningstar}.
The data have been transformed using Eq.~\eqref{V-transform}.
The error bars on the potentials are the statistical errors only.
The curve is the best fit to a quadratic  polynomial in $r$.
}
\label{fig:DeltaVPig}
\end{figure}

Since the $\Pi_g$ potentials associated with the $1^{--}$ gluelump
and the  $2^{--}$ gluelump
have the same B\nobreakdash-O quantum numbers, there can be mixing between the potentials.
The mixing ensures that there can be no crossing between the  adiabatic potentials $\Pi_g$ and  $\Pi_g^\prime$.
The $\Pi_g$ and $\Pi_g^\prime$ potentials were  both calculated 
using the 4 ensembles in Ref.~\cite{Morningstar}.
The difference between the $\Pi_g^\prime$ and $\Pi_g$ potentials in the small-$r$ region 
after the transformation in Eq.~\eqref{V-transform} is shown in Fig.~\ref{fig:DeltaVPig}.
The difference decreases to a minimum  and then it increases.
This is the classic behavior of a narrow avoided crossing.
The difference can be fit to a quadratic polynomial in $r$:
\beq
V_{\Pi_g^\prime}(r) - V_{\Pi_g}(r)  \approx E_x + D_x\,  (r - r_x)^2,
\label{DeltaVPig}
\eeq
with the parameters 
\beq
r_x/r_0 = 0.728(37), \qquad r_0 E_x  = 0.103(24) \qquad  r_0^3 D_x = 0.88(16).
\label{DeltaVPig-params}
\eeq
If $r_0 = 0.5$~fm, the avoided crossing is at  0.36~fm and the energy gap is  41~MeV.

\begin{figure}[t]
\centerline{ \includegraphics*[width=12cm,clip=true]{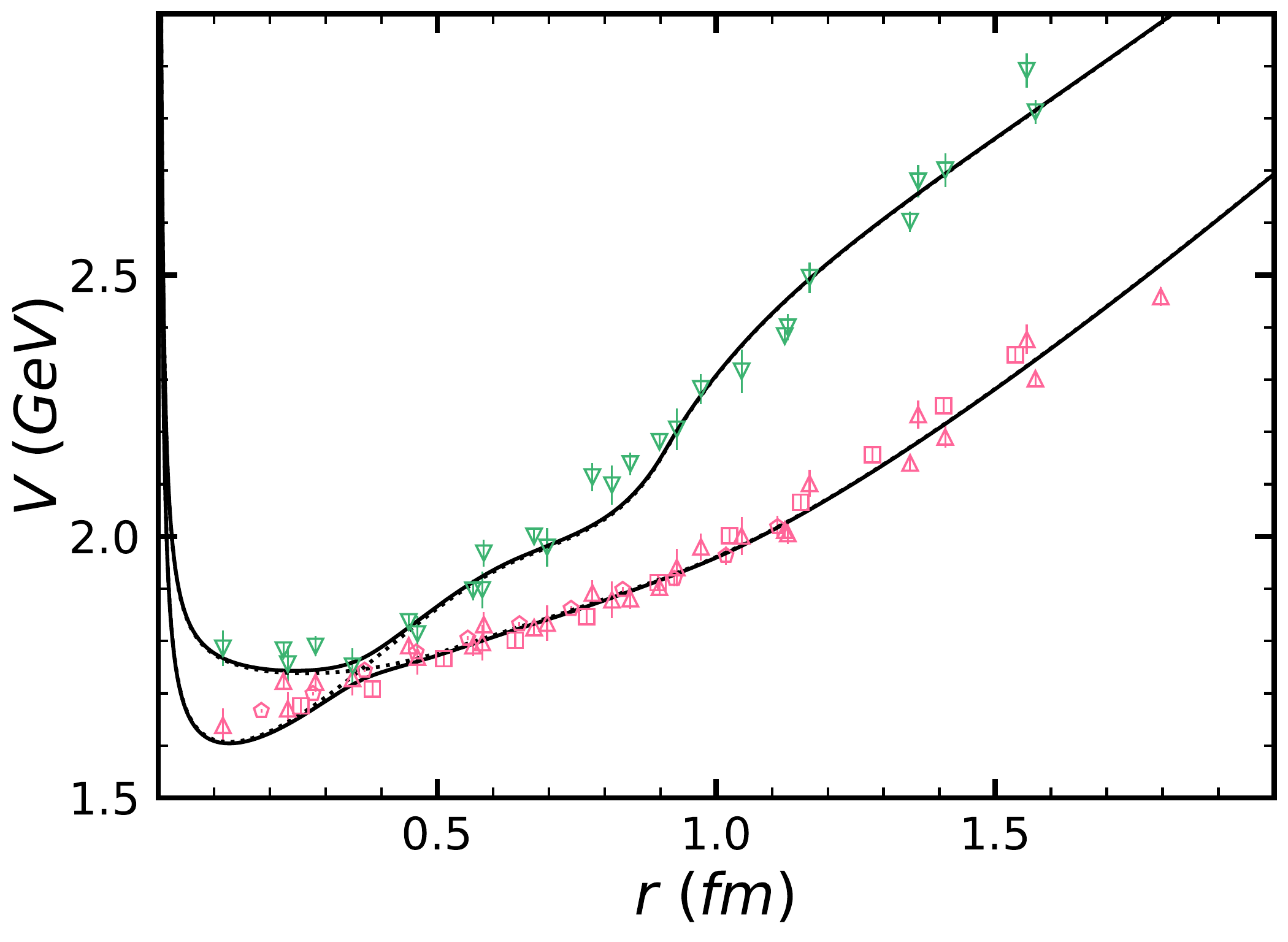}}
\caption{
The $\Pi_g$ and $\Pi_g^\prime$ potentials  in pure $SU(3)$ gauge theory.
The $\Pi_g$ data points are from the 4 ensembles in Ref.~\cite{Morningstar} (up triangles),
the HYP2 ensemble in Ref.~\cite{Capitani:2018rox}  (pentagons),
and the $S_4$  ensemble in Ref.~\cite{Sharifian:2023idc} (squares).
The $\Pi_g^\prime$ data points are from the 4 ensembles in Ref.~\cite{Morningstar} (down triangles).
The data from  Refs.~\cite{Morningstar,Sharifian:2023idc}
have been transformed using Eq.~\eqref{V-transform}.
Physical units  are obtained by setting $r_0 = 0.5$~fm.
The error bars on the potentials are the statistical errors only.
The solid curves are the adiabatic $\Pi_g$ and $\Pi_g^\prime$ potentials  in Eqs.~\eqref{VPiggp} with an avoided crossing.
The dotted curves are the diabatic $\Pi_{g1}$ and $\Pi_{g2}$ potentials.
}
\label{fig:VPig}
\end{figure}

The data for the $\Pi_g$  potentials in Fig.~\ref{fig:VSigmaPig-fit} 
and the $\Pi_g^\prime$ potentials in Fig.~\ref{fig:VDeltaPiSigmag-fit} are shown in Fig.~\ref{fig:VPig}.
 The potential $V_{\Pi_g^\prime}(r)$ above $r_x$ appears to be a continuation of $V_{\Pi_g}(r)$ from below $r_x$,
 while the potential $V_{\Pi_g}(r)$ above $r_x$ appears to be a continuation of $V_{\Pi_g^\prime}(r)$ from below $r_x$.
The avoided crossing between the $\Pi_g^\prime$ and $\Pi_g$ potentials can be explained
 by assuming that the $\Pi_{g1}$ potential associated with the $1^{--}$ gluelump approaches the $N=4$ string excitation at large $r$
while  the $\Pi_{g2}$ potential associated with the $2^{--}$ gluelump approaches the $N=2$ string excitation at large $r$.
In the absence of mixing, they would have to cross at some intermediate $r$.
The crossing appears in the small-$r$ region because the energy difference between the $2^{--}$ and $1^{--}$ gluelumps is small.
The gluelump energy difference calculated in Ref.~\cite{Herr:2023xwg} is 181(38)~MeV.

To parametrize the $\Pi_g^\prime$ and $\Pi_g$ potentials with an avoided crossing, 
we first parametrize the $\Pi_{g1}$ and  $\Pi_{g2}$ potentials in the absence of mixing.
The difference between the $\Pi_{g1}$ and $\Pi_u$ potentials 
can be parametrized as in Eq.~\eqref{VLambdaeta} with $J^{PC} = 1^{--}$ and $N = 4$.
The difference between the $\Pi_{g2}$ and $\Pi_u$ potentials can be parametrized as in
Eq.~\eqref{VLambdaeta} with $J^{PC} = 2^{--}$ and $N =2$.
The adjustable parameters are the coefficients $\Delta  A_{\Pi_{g1}}$, $\Delta  B_{\Pi_{g1}}$, 
$\Delta  A_{\Pi_{g2}}$, and $\Delta  B_{\Pi_{g2}}$ and  two matching radii $r_{\Pi_{g1}}$ and $r_{\Pi_{g2}}$.

The potentials $V_{\Pi_{g1}}(r)$ and $V_{\Pi_{g2}}(r)$ 
can be interpreted as diabatic potentials that are the diagonal entries of a $2\times 2$ diabatic potential matrix.
Since the avoided crossing is narrow, the off-diagonal entries of the diabatic potential matrix
can be approximated by a constant $E_x/2$. 
The adiabatic $\Pi_g$ and $\Pi_g^\prime$  potentials are the eigenvalues of this matrix:
\begin{subequations}
\bqa
V_{\Pi_g}(r) &=&  \tfrac12 \big[ V_{\Pi_{g1}}(r)  + V_{\Pi_{g2}}(r) \big]
-  \tfrac12 \sqrt{[ V_{\Pi_{g1}}(r) - V_{\Pi_{g2}}(r) ]^2 + E_x^2} \,,
\label{VPig}%
\\
V_{\Pi_g^\prime}(r)  &=&   \tfrac12 \big[ V_{\Pi_{g1}}(r)  + V_{\Pi_{g2}}(r) \big]
+  \tfrac12 \sqrt{[ V_{\Pi_{g1}}(r) - V_{\Pi_{g2}}(r) ]^2 + E_x^2} \,.
\label{VPigp}%
\eqa
\label{VPiggp}%
\end{subequations}
The minimum of the difference between the potentials is $E_x$ 
and the crossing radius $r_x$ satisfies $V_{\Pi_{g1}}(r_x) = V_{\Pi_{g2}}(r_x)$.
Since the avoided crossing is in the small-$r$ region,
the equation $V_{\Pi_{g1}}(r_x) = V_{\Pi_{g2}}(r_x)$ imposes a constraint
on the  coefficients  
in the expressions for $V_{\Pi_{g1}}(r)$ and $V_{\Pi_{g2}}(r)$ in Eq.~\eqref{VLambdaeta}:
\beq
(E_{2^{--}} - E_{1^{--}}) + (\Delta A_{\Pi_{g2}} -  \Delta A_{\Pi_{g1}})\,  r_x^2  
+ (\Delta B_{\Pi_{g2}} -  \Delta B_{\Pi_{g1}})\,  r_x^4 + (\Delta C_{\Pi_{g2}} -  \Delta C_{\Pi_{g1}})\,  r_x^6 = 0.
\label{DeltaV-rx}
\eeq
We choose the dimensionless energy difference between the $2^{--}$ and $1^{--}$ gluelumps 
to be the central value from  Ref.~\cite{Herr:2023xwg}:
$r_0(E_{2^{--}} - E_{1^{--}}) = 0.459$. 
We choose the avoided-crossing parameters to be the central values in Eq.~\eqref{DeltaVPig-params}:
$r_x/r_0 = 0.728$ and $r_0 E_x =  0.103$.
The coefficient $\Delta B_{\Pi_{g2}}$ can be eliminated using Eq.~\eqref{DeltaV-rx}.
The  remaining adjustable parameters in the potentials in Eqs.~\eqref{VPiggp} are 
$\Delta A_{\Pi_{g1}}$, $\Delta B_{\Pi_{g1}}$, 
$\Delta A_{\Pi_{g2}}$, $r_{\Pi_{g1}}$, and $r_{\Pi_{g2}}$.
We determine the 5 adjustable parameters 
by minimizing the error-weighted $\chi^2$ for the 50 $\Pi_g$ data points and the 28 $\Pi_g^\prime$ data points.
The fitted dimensionless parameters are given in Table~\ref{tab:Hybrid-params}.
Our fits for the adiabatic $\Pi_g$ and $\Pi_g^\prime$ potentials are shown in Fig.~\ref{fig:VPig}.


\section{Summary}
\label{sec:Summary}

We have presented parameterizations of 8 of the lowest B\nobreakdash-O potentials for pure $SU(3)$ gauge theory.
At small $r$, the 7 hybrid potentials constitute the 3 lowest multiplets of hybrid B\nobreakdash-O potentials,
which are associated with the $1^{-+}$, $1^{--}$, and $2^{--}$ gluelumps.
At large $r$, the 7 hybrid potentials constitute the 2 lowest multiplets of hybrid B\nobreakdash-O potentials
with string excitation levels $N=1$ and 2, one potential with $N=3,$ and two potentials with $N=4$. 

Our parameterization for a potential with B\nobreakdash-O quantum numbers $\Lambda_\eta^\epsilon$
consists of different analytic expressions below and above a matching radius $r_{\Lambda_\eta^\epsilon}$, with coefficients that are constrained by continuity and smoothness at the matching radius.
The parameterizations  below $r_{\Lambda_\eta^\epsilon}$ build in the constraints 
from BOEFT on the small-$r$ behavior of the $\Sigma_g^+$ potential in Eq.~\eqref{VSigmag+-smallr} 
and the hybrid potentials in Eq.~\eqref{VLambda-smallr}.
The parameterizations  above $r_{\Lambda_\eta^\epsilon}$ build in the stringy constraints on the large-$r$ behavior 
 in Eq.~\eqref{V-larger}.

Our parameterization for the B\nobreakdash-O potentials for pure $SU(3)$ gauge theory
are chosen so that they can be easily be modified to take into account information on the potentials from QCD with light quarks.
The parameters in the potentials are expressed as dimensionless parameters 
by multiplying them by the appropriate power of the Sommer scale $r_0$.
The dimensionless parameters are given in Tables~\ref{tab:SigmagpPiu-params} and \ref{tab:Hybrid-params}.  Note that our parameterization for the B-O potentials $\Lambda_\eta^\sigma$ can be improved further by including additional  higher order terms below the matching radius $r_{\Lambda^\sigma_\eta}$.

Our potentials above the matching radius are expressed in terms of the string potentials $V_N(r)$ in Eq.~\eqref{V-N},
which depend only on the tension $\sigma$.
If the dimensionless string tension $r_0^2 \sigma$ is calculated accurately using lattice QCD, 
that value of $\sigma$ can be used in the parameterizations.
Alternatively one can use a phenomenological value of $\sigma$ obtained by fitting bottomonium and charmonium energies to the spectra of energy levels in a model for the $\Sigma_g^+$ potential.

Our parameterizations of the hybrid potentials below the matching radius 
include the appropriate gluelump energy $E_{J^{PC}}$ as an additive constant.
The gluelump energy differences $\Delta E_{J^{PC}} = E_{J^{PC}} - E_{1^{-+}}$ were constrained 
to the values calculated using  pure $SU(3)$ gauge theory in Ref.~\cite{Herr:2023xwg}.
The gluelump energy differences have been calculated by Marsh and Lewis 
using lattice QCD with 2+1 flavors of dynamical light quarks and a pion mass of about $3.5\, m_\pi$ \cite{Marsh:2013xsa}.
The ordering in energy of the first few gluelumps is the same as in pure $SU(3)$ gauge theory.
Replacing the gluelump energy differences in our parameterizations by the
values of $\Delta E_{J^{PC}}$ from lattice QCD would take into account some of the important effects of light quarks.

It has always been clear that the light quarks of QCD require the existence of new B\nobreakdash-O potentials
that approach constants at large $r$ equal to static-hadron pair thresholds.
The static hadron can be classified according to their flavor 
with respect to the approximate $SU(3)$ symmetry relating the $u$, $d$, and $s$ quarks.
As pointed out in Refs.~\cite{Braaten:2013boa,Braaten:2014qka}, 
light quarks also require the existence of new B\nobreakdash-O potentials at small $r$.
They follow from the existence of adjoint hadrons, 
which are discrete energy levels of QCD bound to a static color-octet ($\bm{8}$) source.
The adjoint hadrons can be classified according to their $SU(3)$ flavors.
The only information about adjoint hadrons from lattice gauge theory comes from
calculations by Foster and Michael in 1998 \cite{Foster:1998wu}.
They calculated the energies of adjoint mesons using $SU(3)$ gauge theory with valence light quarks.
They found that the energy of adjoint mesons was close to the energy of the ground-state $1^{-+}$ gluelump
in two specific $J^{PC}$ channels:  $1^{--}$ and $0^{-+}$.

As $r \to 0$,  the new potential associated with an adjoint hadron approaches 
the repulsive color-octet potential between a static quark and antiquark plus the energy of the adjoint hadron.
There is as yet no information from lattice QCD about the B\nobreakdash-O potential associated with any adjoint hadron.
One possibility is that as the separation $r$ of  the $\bm{3}$ and $\bm{3}^\ast$ sources increases,
the light quarks and antiquarks in the adjoint hadron arrange themselves into  
a cluster bound to the $\bm{3}$ source  with total color charge $\bm{3}^\ast$
and a cluster bound to the $\bm{3^\ast}$ source  with total color charge  $\bm{3}$.
The clusters with color charge $\bm{3}^\ast$ and $\bm{3}$ would be connected by a gluon flux tube,
so the potential would increase linearly with $r$.
A plausible model for such a B\nobreakdash-O potential  is a potential from pure $SU(3)$ gauge theory
with the gluelump energy replaced by the energy of the adjoint hadron.

Improved models for the B\nobreakdash-O potentials for  QCD based on those  for  pure $SU(3)$ gauge theory should  
 allow the further development of the  B\nobreakdash-O approximation for  QCD. 
 It could lead to progress in understanding the  exotic heavy hadrons.
 However definitive calculations of  the B\nobreakdash-O potentials for  QCD using lattice gauge theory will be essential for 
 quantitative predictions of QCD for exotic heavy hadrons.

\begin{acknowledgments}
This research was supported in part by the Department of Energy under grant DE-FG02-05ER15715,
by the DFG Project-ID 196253076 TRR 110, and by the NSFC through
funds provided to the Sino-German CRC 110 ``Symmetries and the Emergence of Structure in QCD". 
AM and EB acknowledge the DFG cluster of excellence ORIGINS funded by the Deutsche Forschungsgemeinschaft 
under Germany’s Excellence Strategy EXC-2094-390783311. 
AM also acknowledges the ``Strong Interaction at the Frontier of Knowledge: Fundamental Research and Applications” project
funded by the European Union’s Horizon 2020 program under grant 824093.
We thank M.~Wagner for valuable insights into the fitting of B-O potentials.
We thank N.~Brambilla and A.~Vairo for useful discussions. 
This work contributes to the goals of the US DOE ExoHad Topical Collaboration, Contract DE-SC0023598.
\end{acknowledgments}




%


\begin{thebibliography}{0}%
\makeatletter
\providecommand \@ifxundefined [1]{%
 \@ifx{#1\undefined}
}%
\providecommand \@ifnum [1]{%
 \ifnum #1\expandafter \@firstoftwo
 \else \expandafter \@secondoftwo
 \fi
}%
\providecommand \@ifx [1]{%
 \ifx #1\expandafter \@firstoftwo
 \else \expandafter \@secondoftwo
 \fi
}%
\providecommand \natexlab [1]{#1}%
\providecommand \enquote  [1]{``#1''}%
\providecommand \bibnamefont  [1]{#1}%
\providecommand \bibfnamefont [1]{#1}%
\providecommand \citenamefont [1]{#1}%
\providecommand \href@noop [0]{\@secondoftwo}%
\providecommand \href [0]{\begingroup \@sanitize@url \@href}%
\providecommand \@href[1]{\@@startlink{#1}\@@href}%
\providecommand \@@href[1]{\endgroup#1\@@endlink}%
\providecommand \@sanitize@url [0]{\catcode `\\12\catcode `\$12\catcode
  `\&12\catcode `\#12\catcode `\^12\catcode `\_12\catcode `\%12\relax}%
\providecommand \@@startlink[1]{}%
\providecommand \@@endlink[0]{}%
\providecommand \url  [0]{\begingroup\@sanitize@url \@url }%
\providecommand \@url [1]{\endgroup\@href {#1}{\urlprefix }}%
\providecommand \urlprefix  [0]{URL }%
\providecommand \Eprint [0]{\href }%
\providecommand \doibase [0]{https://doi.org/}%
\providecommand \selectlanguage [0]{\@gobble}%
\providecommand \bibinfo  [0]{\@secondoftwo}%
\providecommand \bibfield  [0]{\@secondoftwo}%
\providecommand \translation [1]{[#1]}%
\providecommand \BibitemOpen [0]{}%
\providecommand \bibitemStop [0]{}%
\providecommand \bibitemNoStop [0]{.\EOS\space}%
\providecommand \EOS [0]{\spacefactor3000\relax}%
\providecommand \BibitemShut  [1]{\csname bibitem#1\endcsname}%
\let\auto@bib@innerbib\@empty
\end{thebibliography}%


\begin{thebibliography}{99}

  
\bibitem{Brambilla:2019esw}
N.~Brambilla, S.~Eidelman, C.~Hanhart, A.~Nefediev, C.P.~Shen, C.E.~Thomas, A.~Vairo and C.Z.~Yuan,
The $XYZ$ states: experimental and theoretical status and perspectives,'
Phys.\ Rept.\ \textbf{873}, 1-154 (2020)
[arXiv:1907.07583].

\bibitem{Juge:1999ie}
K.J.~Juge, J.~Kuti and C.~Morningstar,
Ab initio study of hybrid $\bar{b} g b$ mesons,
Phys.\ Rev.\ Lett.\ \textbf{82}, 4400-4403 (1999)
[hep-ph/9902336].

\bibitem{Brambilla:1999xf} 
 N.~Brambilla, A.~Pineda, J.~Soto and A.~Vairo,
Potential NRQCD: an effective theory for heavy quarkonium,
 Nucl.\ Phys.\ B {\bf 566}, 275 (2000)
 [hep-ph/9907240].
  

\bibitem{Berwein:2015vca}
M.~Berwein, N.~Brambilla, J.~Tarr\'us Castell\`a and A.~Vairo,
Quarkonium Hybrids with Nonrelativistic Effective Field Theories,
Phys.\ Rev.\ D \textbf{92},  114019 (2015)
[arXiv:1510.04299].

\bibitem{Oncala:2017hop}
R.~Oncala and J.~Soto,
Heavy Quarkonium Hybrids: Spectrum, Decay and Mixing,
Phys.\ Rev.\ D \textbf{96}, 014004 (2017)
[arXiv:1702.03900].

\bibitem{Brambilla:2017uyf}
N.~Brambilla, G.~Krein, J.~Tarr\'us Castell\`a and A.~Vairo,
Born-Oppenheimer approximation in an effective field theory language,
Phys.\ Rev.\ D \textbf{97}, 016016 (2018)
[arXiv:1707.09647].

\bibitem{Soto:2020xpm}
J.~Soto and J.~Tarr\'us Castell\`a,
Nonrelativistic effective field theory for heavy exotic hadrons,
Phys.\ Rev.\ D \textbf{102},  014012 (2020)
[arXiv:2005.00552].

\bibitem{Berwein:2024ztx}
M.~Berwein, N.~Brambilla, A.~Mohapatra and A.~Vairo,
One Born$-$Oppenheimer Effective Theory to rule them all: hybrids, tetraquarks, pentaquarks, doubly heavy baryons and quarkonium,
[arXiv:2408.04719].
  
\bibitem{Brambilla:2018pyn}
N.~Brambilla, W.K.~Lai, J.~Segovia, J.~Tarr\'us Castell\`a and A.~Vairo,
Spin structure of heavy-quark hybrids,
Phys.\ Rev.\ D \textbf{99},  014017 (2019)
[arXiv:1805.07713].

\bibitem{Brambilla:2019jfi}
N.~Brambilla, W.K.~Lai, J.~Segovia and J.~Tarr\'us Castell\`a,
QCD spin effects in the heavy hybrid potentials and spectra,
Phys.\ Rev.\ D \textbf{101}, 054040 (2020)
[arXiv:1908.11699].

\bibitem{Soto:2023lbh}
J.~Soto and S.T.~Valls,
Hyperfine splittings of heavy quarkonium hybrids,
Phys.\ Rev.\ D \textbf{108}, 014025 (2023)
[arXiv:2302.01765].

\bibitem{TarrusCastella:2021pld}
J.~Tarr\'us Castell\`a and E.~Passemar,
Exotic to standard bottomonium transitions,
Phys.\ Rev.\ D \textbf{104}, 034019 (2021)
[arXiv:2104.03975].

\bibitem{Brambilla:2022hhi}
N.~Brambilla, W.K.~Lai, A.~Mohapatra and A.~Vairo,
Heavy hybrid decays to quarkonia,
Phys.\ Rev.\ D \textbf{107}, 054034 (2023)
[arXiv:2212.09187].

\bibitem{Soto:2020pfa}
J.~Soto and J.~Tarr\'us Castell\`a,
Effective field theory for double heavy baryons at strong coupling,
Phys.\ Rev.\ D \textbf{102}, 014013 (2020)
[arXiv:2005.00551].

\bibitem{TarrusCastella:2022rxb}
J.~Tarr\'us Castell\`a,
Heavy meson thresholds in Born-Oppenheimer effective field theory,
Phys.\ Rev.\ D \textbf{106}, 094020 (2022)
[arXiv:2207.09365].

\bibitem{Juge:2002br}
K.~Juge, J.~Kuti and C.~Morningstar,
Fine structure of the QCD string spectrum,
Phys.\ Rev.\ Lett.\ \textbf{90}, 161601 (2003)
[arXiv:hep-lat/0207004].  

\bibitem{Morningstar}
\verb|https://www.andrew.cmu.edu/user/cmorning/static_potentials/SU3_4D/greet.html|.
  
\bibitem{Capitani:2018rox}
S.~Capitani, O.~Philipsen, C.~Reisinger, C.~Riehl and M.~Wagner,
Precision computation of hybrid static potentials in $SU(3)$ lattice gauge theory,
Phys.\ Rev.\ D \textbf{99}, 034502 (2019)
[arXiv:1811.11046].

\bibitem{Schlosser:2021wnr}
C.~Schlosser and M.~Wagner,
Hybrid static potentials in SU(3) lattice gauge theory at small quark-antiquark separations,
Phys.\ Rev.\ D \textbf{105}, 054503 (2022)
[arXiv:2111.00741].

\bibitem{Bicudo:2021tsc}
P.~Bicudo, N.~Cardoso and A.~Sharifian,
Spectrum of very excited $\Sigma_g^+$ flux tubes in $SU(3)$ gauge theory,
Phys.\ Rev.\ D \textbf{104}, 054512 (2021)
[arXiv:2105.12159].

\bibitem{Sharifian:2023idc}
A.~Sharifian, N.~Cardoso and P.~Bicudo,
Eight very excited flux tube spectra and possible axions in $SU(3)$ lattice gauge theory,
Phys.\ Rev.\ D \textbf{107}, 114507 (2023)
[arXiv:2303.15152].

\bibitem{Bali:2005fu} 
  G.S.~Bali {\it et al.}  [SESAM Collaboration],
Observation of string breaking in QCD,
  Phys.\ Rev.\ D {\bf 71}, 114513 (2005)
  [hep-lat/0505012].

\bibitem{Bulava:2019iut}
J.~Bulava, B.~H\"orz, F.~Knechtli, V.~Koch, G.~Moir, C.~Morningstar and M.~Peardon,
String breaking by light and strange quarks in QCD,
Phys.\ Lett.\ B \textbf{793}, 493-498 (2019)
[arXiv:1902.04006].

\bibitem{Bulava:2024jpj}
J.~Bulava, F.~Knechtli, V.~Koch, C.~Morningstar and M.~Peardon,
The quark-mass dependence of the potential energy between static colour sources in the QCD vacuum with light and strange quarks,
[arXiv:2403.00754 [hep-lat]].

\bibitem{Bali:2000vr}
G.S.~Bali \textit{et al.} [TXL and T(X)L],
Static potentials and glueball masses from QCD simulations with Wilson sea quarks,
Phys.\ Rev.\ D \textbf{62}, 054503 (2000)
[:hep-lat/0003012].

\bibitem{Hollwieser:2023bud}
R.~H\"ollwieser, F.~Knechtli, T.~Korzec, M.~Peardon and J.A.~Urrea-Ni\~no,
Hybrid static potentials from Laplacian Eigenmodes,
[arXiv:2401.09453 [hep-lat]].

\bibitem{Braaten:2013boa}
E.~Braaten,
How the $Z_c$(3900) Reveals the Spectra of Quarkonium Hybrid and Tetraquark Mesons,
Phys.\ Rev.\ Lett.\ \textbf{111}, 162003 (2013)
[arXiv:1305.6905].

\bibitem{Braaten:2014qka} 
 E.~Braaten, C.~Langmack and D.H.~Smith,
Born-Oppenheimer approximation for the $XYZ$ mesons,
Phys.\ Rev.\ D {\bf 90}, 014044 (2014)
  [arXiv:1402.0438].

\bibitem{Prelovsek:2019ywc}
S.~Prelovsek, H.~Bahtiyar and J.~Petkovic,
$Z_b$ tetraquark channel from lattice QCD and Born-Oppenheimer approximation,
Phys.\ Lett.\ B \textbf{805}, 135467 (2020)
[arXiv:1912.02656].

\bibitem{Sadl:2021bme}
M.~Sadl and S.~Prelovsek,
Tetraquark systems $\bar bb \bar du$ in the static limit and lattice QCD,
Phys.\ Rev.\ D \textbf{104},  114503 (2021)
[arXiv:2109.08560].

\bibitem{Juge:2003ge}
K.J.~Juge, J.~Kuti and C.~Morningstar,
Excitations of the static quark anti-quark system in several gauge theories,
contribution to Confinement 2003
[arXiv:hep-lat/0312019].

\bibitem{Luscher:2004ib}
M.~Luscher and P.~Weisz,
String excitation energies in $SU(N)$ gauge theories beyond the free-string approximation,
JHEP \textbf{07}, 014 (2004)
[hep-th/0406205].

\bibitem{Braaten:1986bz}
E.~Braaten, R.D.~Pisarski and S.-M.~Tse,
The Static Potential for Smooth Strings,
Phys.\ Rev.\ Lett.\ \textbf{58}, 93 (1987).

\bibitem{Braaten:1987gq}
E.~Braaten and S.-M.~Tse,
The Static Potential for Smooth Strings in the Large $D$ Limit,
Phys.\ Rev.\ D \textbf{36}, 3102 (1987).

\bibitem{Aharony:2010cx}
O.~Aharony and M.~Field,
On the effective theory of long open strings,
JHEP \textbf{01}, 065 (2011)
[arXiv:1008.2636].

\bibitem{Foster:1998wu}
M.~Foster \textit{et al.} [UKQCD],
Hadrons with a heavy color adjoint particle,
Phys.\ Rev.\ D \textbf{59}, 094509 (1999)
[hep-lat/9811010].

\bibitem{Herr:2023xwg}
J.~Herr, C.~Schlosser and M.~Wagner,
Gluelump masses and mass splittings from SU(3) lattice gauge theory,
Phys. Rev. D \textbf{109}, 034516 (2024)
[arXiv:2306.09902].
\\UPDATED

\bibitem{Sommer:1993ce}
R.~Sommer,
A New way to set the energy scale in lattice gauge theories and its applications to the static force and $\alpha_s$ in $SU(2)$ Yang-Mills theory,
Nucl.\ Phys.\ B \textbf{411}, 839-854 (1994)
[arXiv:hep-lat/9310022].


\bibitem{Karbstein:2018mzo}
F.~Karbstein, M.~Wagner and M.~Weber,
Determination of $\Lambda_{\overline{\textrm{MS}}}^{(n_f=2)}$ and analytic parametrization of the static quark-antiquark potential,
Phys.\ Rev.\ D \textbf{98}, no.11, 114506 (2018)
[arXiv:1804.10909].

\bibitem{Pineda:2019mhw}
A.~Pineda and J.~Tarr\'us Castell\'a,
Novel implementation of the multipole expansion to quarkonium hadronic transitions,
Phys.\ Rev.\ D \textbf{100}, 054021 (2019)
[arXiv:1905.03794].

\bibitem{Marsh:2013xsa} 
 K.~Marsh and R.~Lewis,
A lattice QCD study of generalized gluelumps,
  Phys.\ Rev.\ D {\bf 89}, 014502 (2014)
  [arXiv:1309.1627].


  
\end{thebibliography}
\end{document}